\input{epsf}

\documentclass[journal]{IEEEtran}

\makeatletter
\def\ps@headings{%
\def\@oddhead{\mbox{}\scriptsize\rightmark \hfil \thepage}%
\def\@evenhead{\scriptsize\thepage \hfil \leftmark\mbox{}}%
\def\@oddfoot{}%
\def\@evenfoot{}}

\makeatother
\usepackage{epsf}
\usepackage{graphicx}
\usepackage[cmex10]{amsmath}
\usepackage{amsthm}
\usepackage{amssymb}
\usepackage{epsfig,latexsym,amsmath,epsf,amssymb,amsfonts}
\usepackage{algorithm, algcompatible}

\usepackage{placeins}
\usepackage{url}
\usepackage{cite}
\usepackage{comment}
\usepackage{subfigure}
\usepackage{lipsum, mathtools, cuted}
\usepackage{stfloats}
\usepackage{physics}
\usepackage{color}

\usepackage[top=0.75in, bottom=1in, left=0.625in, right=0.625in]{geometry}
\usepackage{bbm}

\renewcommand{\theequation}{\arabic{equation}}
\newtheorem{Definition}{\hskip 0pt Definition}%[section]
\newtheorem{Theorem}{\hskip 0pt Theorem}%[section]
\newtheorem{Corollary}{\hskip 0pt Corollary}%[section]
\newtheorem{Remark}{\hskip 0pt Remark}%[section]
\usepackage{tikz}
\usetikzlibrary{shapes.geometric, arrows, calc}

\usepackage{newfloat}
\DeclareFloatingEnvironment{Algorithm}
\newcommand{\vp}{\mbox{${\bf p}$}}

\newcommand{\gd}{\delta}
\newcommand{\gre}{\varepsilon}

\newcommand{\gD}{\Delta}

\newtheorem{prop}{Proposition}
\newtheorem{coro}{Corollary}

\newcommand{\bea}{\begin{array}}
\newcommand{\ena}{\end{array}}
\newcommand{\bds}{\begin {description}}
\newcommand{\eds}{\end {description}}
\newcommand{\bdf}{\begin{definition}}
\newcommand{\blm}{\begin{lemma}}
\newcommand{\edf}{\end{definition}}
\newcommand{\elm}{\end{lemma}}
\newcommand{\bthm}{\begin{theorem}}
\newcommand{\ethm}{\end{theorem}}
\newcommand{\bprp}{\begin{prop}}
\newcommand{\eprp}{\end{prop}}
\newcommand{\bcl}{\begin{claim}}
\newcommand{\ecl}{\end{claim}}
\newcommand{\bcr}{\begin{coro}}
\newcommand{\ecr}{\end{coro}}
\newcommand{\bquest}{\begin{question}}
\newcommand{\equest}{\end{question}}

\begin{document}

% paper title
%\title{Decentralized Power Allocation and Beamforming Using Non-Convex Nash Game for Energy-Aware mmWave Networks}
\title{Non-Convex Generalized Nash Games for Energy Efficient Power Allocation and Beamforming in mmWave Networks}

\author{
  \IEEEauthorblockN{Wenbo Wang~\IEEEmembership{Member,~IEEE}
  and
  Amir Leshem~\IEEEmembership{Fellow,~IEEE}
  }
  \thanks{This paper has been accepted by IEEE Transactions on Signal Processing. Part of this paper has been presented in~\cite{wenbo_icc_2022} and~\cite{Wang2206:Power}.}
  \thanks{Wenbo Wang and Amir Leshem are with the Faculty of Engineering, Bar-Ilan University, Ramat-Gan, Israel 52900 (email: wangwen@biu.ac.il, leshema@biu.ac.il).}
  \thanks{Copyright~\copyright~2022 IEEE. Personal use of this material is permitted. Permission from IEEE must be obtained for all other uses, in any current or future media, including reprinting/republishing this material for advertising or promotional purposes, creating new collective works, for resale or redistribution to servers or lists, or reuse of any copyrighted component of this work in other works by sending a request to pubs-permissions@ieee.org.}
}

\maketitle

\begin{abstract}
Network management is a fundamental ingredient for efficient operation of wireless networks. With increasing bandwidth, number of antennas and number of users, the amount of information required for network management increases significantly. Therefore, distributed network management is a key to efficient operation of future networks. This paper focuses on the problem of distributed joint beamforming control and power allocation in ad-hoc mmWave networks. Over the shared spectrum, a number of multi-input-multi-output links attempt to minimize their supply power by simultaneously finding the locally optimal power allocation and beamformers in a self-organized manner. Our design considers a family of non-convex quality-of-service constraint and utility functions characterized by monotonicity in the strategies of the various users. We propose a two-stage, decentralized optimization scheme, where the adaptation of power levels and beamformer coefficients are iteratively performed  by each link. We first prove that given a set of receive beamformers, the power allocation stage converges to an optimal generalized Nash equilibrium of the generalized power allocation game. Then we prove that iterative minimum-mean-square-error adaptation of the receive beamformer results in an overall converging scheme. Several transmit beamforming schemes requiring different levels of information exchange are also compared in the proposed allocation framework. Our simulation results show that allowing each link to optimize its transmit filters using the direct channel results in a near optimum performance with very low computational complexity, even though the problem is highly non-convex.

\end{abstract}

\section{Introduction}\label{Sec_introduction}
The development of 5G technologies has pushed the spectrum utilization of wireless networks towards the millimeter wave (mmWave) regime~\cite{6736752}. With the help of multiple antennas in compensating for excessive path losses, mmWave communication is considered to be able to improve network throughput by orders of magnitude. In a typical mmWave deployment scenario for urban outdoor coverage, distributed large antenna arrays in the form of various types of Base Stations (BSs) and small-cell (e.g., femtocell) Access Points (APs) are densely deployed. However, this naturally leads to a prohibitive capital expenditure for the conventional backhaul infrastructure, which relies on fiber optical cables for establishing connections between the BSs/APs and the core network~\cite{7400949}. Since it is possible to construct high-rate Multiple-Input-Multiple-Output (MIMO) links with a dominant Line-of-Sight (LoS) component over short distances among the BSs/APs, the wireless backhaul for cellular systems becomes a highly desirable and cost-effective solution. Regarding the recent progress of related research and standardization for wireless backhauls in the context of 5G, see~\cite{9040265} for an excellent overview.

Nevertheless, with the wireless backhaul over mmWave spectrum, the networks confront the problems of interference management between different (e.g., inter-cell) data streams and a significant increase in the power consumption of the dedicated Radio Frequency (RF)-chains. From an implementation point of view, current interference and power management typically relies on a centralized controller, such as the cloudified baseband unit pools or cluster-based BS controllers. In this centralized framework, the logical links to the RF modules are established with every BS/AP. Then, sufficient information of every link, e.g., the Channel State Information (CSI) and the channel quality indicators, can be collected from each BS/AP for physical-layer coordination (e.g., zero-forcing beamforming of multiple links). This incurs prohibitive overhead due to the required amount of data feedback/exchange. Meanwhile, due to the possible ad-hoc characteristics of small cells, some APs may even face a limit on the way that information exchange takes place both across tiers and among cells~\cite{6990372}. For this reason, a fully distributed link parameter control mechanism for wireless backhaul, where each BS/AP is capable of adapting to the surrounding environment is highly desirable.

To that end, we study the problem of distributed energy-aware power allocation and beamformer design in mmWave communication with ad-hoc links (i.e., wireless backhaul). First, we study the problem of distributed power allocation under a family of non-convex Quality of Service (QoS) constraints. We propose a generalized Nash game formulation of the problem and provide a decentralized solution which minimizes the total supply power under rate and QoS constraints. Our main theoretical result proves that for a certain family of monotonic non-convex generalized games, asynchronous best response strategy is able to converge efficiently to the socially optimum power allocation. Based on this theoretical result, we prove that an iterative solution of the power allocation and beamformer design converges to an optimum (if feasible) and also identifies in-feasibility when the required set of QoS is unattainable. The solution is featured by its high computational efficiency as well as low coordination overhead.

Numerical simulations demonstrate the efficiency of the proposed solution in terms of both the network performance and computational efficiency through comparison with the state-of-the-art centralized precoding-decoding vector adaptation schemes, including the coordinated zero-forcing approaches and Minimum Mean Squared Error (MMSE) interference suppression approaches.

\subsection{Related Works and Contribution}\label{Sec:Survey}
It is well-known that the problem of joint power and spectrum management with single antenna over multiple links is NP-hard~\cite{4453890}. When it comes to the multi-antenna scenarios, early studies on the joint beamforming and power allocation problems for multiple users/links usually consider optimizing the sum-rate~\cite{4509444} or the minimum achievable throughput (i.e., the fairness utility)~\cite{1262126} across the links. In these cases, the formulated network optimization problems are non-convex. Thus, they have to be transformed, with certain relaxation, into computationally tractable forms such as semidefinite programming problems~\cite{6760591} or second-order cone programming problems~\cite{5463229}. Then, an efficient allocation scheme can be developed. Otherwise, decomposition of the global allocation problem into multiple stages, each targeting at solving for a sub-set of optimization variables, is preferred, especially when the target optimization problem involves more than one set of parameters (e.g., joint Tx/Rx filter design)~\cite{9475063, 8048676}. When the problem is limited to the single-BS-multi-user cases (e.g.,~\cite{4509444}), it is convenient to assume that the information of all relevant channels is known at the single BS. Alternatively, the full CSI of the channels associated with the multiple transmitter-receiver pairs has to be obtained by a central coordinator in advance. With such assumptions, the transformed problems can be properly solved in a one-shot optimization manner. %Furthermore, for interference optimization problems involving more than one set of parameters (e.g., joint Tx/Rx filter design),  iterative procedures based on multiple stages of sub-optimization problems are widely adopted~\cite{6760591, 8048676}.

%the ``monotonic program'' discussion is added here
Besides the single-objective optimization-based formulation, Multiple-Objective Optimization (MOO) is also introduced, from a centralized perspective, for deriving proper transmit strategies of multiple links~\cite{bjornson2013optimal}. Compared with the aforementioned studies, MOO-based formulation is able to address conflicting objectives of local links, which are caused by the inter-link interference or power constraint coupling. The solution to an MOO problem is typically provided by searching for the Pareto optimal points in the achievable performance region of the network~\cite{branke2008multiobjective}. In addition, by introducing the subjective system utility functions that are monotonic and Lipschitz continuous, the MOO-based formulation can be further converted (also known as scalarization) to single-objective optimization problems. Based on the analysis of Pareto boundaries~\cite{bjornson2013optimal}, a number of non-convex scenarios can be effectively addressed through certain bounding-and-search procedures (e.g.,~\cite{6119237}) in the framework of monotonic optimization.

When ad-hoc links are considered, a number of studies have attempted to tackle the joint allocation problem over the MIMO interference channel from the perspective of non-cooperative games~\cite{8031014, 8756054}. In order to obtain the Pareto-preferred equilibria, externalities (e.g., pricing) have also been introduced~\cite{6502486}. From the point of view of a single link/user, the solution to equilibrium searching in the formulated allocation game is usually built upon a local (convex) optimization problem with a number of constraints depending solely on the local power-selection/beamforming strategy (see~\cite{6502486} for an example). However, when the strategy space of one link is determined by the joint adversary strategy of the other links, theoretical tools such as generalized Nash Equilibrium (NE) and quasi-variational inequality are needed. To date, most studies in this category of problems are either confined to the single-antenna scenarios (e.g.,~\cite{7145476}) or developed based on the strong assumptions of uplink-downlink duality or time-division duplexing under special topology/protocol requirements of the network (e.g.,~\cite{1621177}).

Compared with these works, especially those employing game theoretic tools for joint power and beamforming control~\cite{6502486, 1621177, 5740995}, our study does not rely on specific assumptions such as a single receive antenna~\cite{5740995}, a special network topology~\cite{1621177} or non-coupling strategy constraints~\cite{6502486}. Compared with the work that also develops network control schemes using generalized Nash games~\cite{5740995}, our proposed management scheme does not impose special prerequisite of the structure of the game, such as the structure of the $M$-matrix in the best-response functions for identifying the existence of NE. Nor do we require the estimation and knowledge of the interfering channels. Our proposed iterative algorithm is purely decentralized and provably converges to the near-optimal power allocation and receive beamformer. Moreover, it only requires the standard interference measurement and channel estimation over each link. The proposed algorithm is implementation-friendly and our simulation results show that the proposed algorithm is able to quickly converge to near optimal allocation and beamformers, when compared with the state-of-the-art joint beamforming and power allocation algorithms for multiple links/cells with full coordination. Finally, we would like to mention, that in \cite{wenbo_icc_2022} we used the proposed monotonic game framework to optimize energy aware ALOHA networks. The paper did not include proofs of the main theorems here (except a brief outline of the proof of Theorem 1), and does not discuss the problem of beamforming or power allocation, but deals with distributed access probability allocation to the different devices.

The main contributions of our study are as follows:
\begin{enumerate}
\item We provide a novel existence theorem for the family of monotonic generalized Nash games with possibly non-convex constraint and cost functions. Moreover, when a feasible solution exists, we provide a finite-time algorithm which converges to the $\varepsilon$-Nash equilibrium.
It is worth noting that the analytical framework based on the proposed generalized Nash games can be applied to a variety of different decentralized allocation problems, ranging from the power allocation problem in this paper to the random access problem over shared channels.
\item We propose a fully decentralized two-stage, power-efficient solution to the problem of joint beamforming and power control in ad-hoc mmWave networks. The proposed protocol is fully distributed, with each link operating independently, requiring no inter-link information exchange.  It has significantly less signaling and estimation overhead required when compared to the solutions of the problem based on explicit inter-link coordination.
\item We prove that the two-stage power allocation and beamformer design is optimal for any given set of transmit beamformers. Moreover, we demonstrate through simulations, that limiting the transmit beamforming filters to be matched to the direct MIMO channel is near optimal, and this leads to significant computational and messaging savings.
\end{enumerate}

The rest of the paper is organized as follows. Section~\ref{Sec:Model} introduces a generalized model of energy efficiency-based joint power allocation and beamforming control in mmWave MIMO. Section~\ref{Sec:Dec_approach} proposes a decentralized approach for joint power and beamforming control based on a two-stage, iterative strategy-searching mechanism. The simulation results of the achievable energy efficiency and network capacity are reported in Section~\ref{Sec:Simulation}. Section~\ref{Sec:Conclusion} concludes the paper.

\section{Network Model and Problem Formulation}\label{Sec:Model}
\subsection{Network Model}
We consider the scenario of wireless backhaul communication in a fixed-topology network over the mmWave band, where $N$ pairs of Serving Stations (SS) transmit to their Destination Stations (DS) simultaneously. Each SS (DS) is equipped with $K$ ($L$) transmit (receive) chains. At a given time instance, SS $n$ transmits a single symbol $s_n$, precoded using a $K$-dimensional complex precoding vector $\mathbf{w}_n$. Without loss of generality, each SS $n$ employs a normalized $K$-dimensional complex precoding vector $\mathbf{w}_n$. Let $\mathbf{x}_n=\mathbf{w}_ns_n$ be the precoded symbol ($\Vert \mathbf{x}_n \Vert^2=1$). Then, for each link $n=1,\ldots, N$, the received signal at the target DS over the SS-DS link $n$ can be expressed as
\begin{equation}
  \label{eq_received_signal}
  \mathbf{y}_n = \sqrt{P_n}\mathbf{H}_{n,n}\mathbf{x}_n + \sum_{i\ne n}\sqrt{P_i}\mathbf{H}_{i,n}\mathbf{x}_i + \boldsymbol{\eta}_n,
\end{equation}
where $P_n$ is the total transmit power of the active antennas on SS $n$, $\mathbf{H}_{i,n}$ is the $L\!\times\! K$ complex channel matrix between SS $i$ and DS $n$, and $\boldsymbol{\eta}_n$ is the $L$-dimensional complex-value i.i.d. additive Gaussian white noise with variance $\sigma_n^2$, $\boldsymbol{\eta}_n\sim\mathcal{CN}(0, \sigma_n^2\mathbf{I})$. Let $\mathbf{u}_n$ denote the $L$-dimensional receive filter vector adopted by DS $n$. We assume that each DS employs the linear receiver as $\hat{s}_n=\mathbf{u}^{\textrm{H}}_n\mathbf{y}_n$. We also assume that the MIMO channels are flat-fading, and each SS adopts the same power over a set of resource blocks, during which the channel gains remain constant. Then, the Signal-to-Interference-plus-Noise-Ratio (SINR) at DS $n$ can be expressed as
\begin{equation}
  \label{eq_SINR}
  \gamma_n = \displaystyle\frac{P_n\Vert \mathbf{u}^{\textrm{H}}_n\mathbf{H}_{n,n}\mathbf{w}_n\Vert^2}{\sum_{i\ne n} P_i\Vert \mathbf{u}^{\textrm{H}}_n\mathbf{H}_{i,n}\mathbf{w}_i\Vert^2 + \sigma_n^2}.
\end{equation}
The general goal of each SS-DS link is to maintain a certain QoS level (e.g., measured based on the SINR in (\ref{eq_SINR})) while minimizing the local power consumption.

\subsection{Problem Formulation}\label{Sec:Formulation}
Our objective is to jointly optimize the beamforming vectors and the transmit powers over each link, to minimize the power consumption while guaranteeing the required QoS level at each DS. Since energy efficiency is measured as the total QoS divided by the total consumed power, minimizing the power when QoS requirements are fixed results in the most energy efficient solution.
From the perspective of social optimality, the problem of joint beamforming-power control and interference coordination can be expressed as follows with a generalized objective function $c_n(\cdot)$ for power consumption:
\allowdisplaybreaks
\begin{subequations}\label{eq_centralized_optimal}
\begin{align}
    (P^*_n, \mathbf{u}^*_n, \mathbf{w}^*_n)_{n=1}^N= \nonumber\\
      \tag{\ref{eq_centralized_optimal}}
    \arg\min\limits_{(P_n, \mathbf{u}_n, \mathbf{w}_n)_{n=1}^N} & \displaystyle\sum_{n=1}^N c_n(P_n, P_{-n}) \\
    \label{eq_central_optimal_a}
    \textrm { s.t.}\qquad & q_n(P_n, \mathbf{u}_n, \mathbf{w}_n, P_{-n}, \mathbf{u}_{-n}, \mathbf{w}_{-n}) \ge \overline{q}_n, \nonumber \\
    & \forall n=1,\ldots, N, \\
    \label{eq_central_optimal_b}
    & \underline{P}_n\le P_n\le \overline{P}_n, \forall n=1,\ldots, N,\\
    \label{eq_central_optimal_c}
    & \Vert \mathbf{w}_n \Vert = 1, \forall n=1,\ldots, N,
\end{align}
\end{subequations}
where $P_{-n}$ and $\mathbf{u}_{-n}$ ($\mathbf{w}_{-n}$) are the joint power allocation and Rx(Tx)-beamforming vectors of the adversary links of link $n$ (namely, the links other than $n$). $\sum_{n=1}^{N}c_n(P_n, P_{-n})$ is the social cost measuring the energy efficiency of the network. We assume that $c_n(\mathbf{p})$ is a monotonically increasing function in $P_n$ for any $1\le n\le N$ with $\mathbf{p}=(P_n, P_{-n})=[P_1,\ldots, P_N]^{\textrm{T}}$. $c_n(\mathbf{p})$ is independent of the transmit/receive (Tx/Rx) beamforming vectors.
$q_n(\mathbf{p}, \mathbf{u}_n, \mathbf{w}_n, \mathbf{u}_{-n}.
\mathbf{w}_{-n})$ is the measured QoS level of link $n$ as a function of the joint power allocation $\mathbf{p}$ and the beamforming vectors of all the links.
For instance, the QoS can be defined, e.g., by the channel capacity, proportional to $q_n(\gamma_n)=\log_2(1+\gamma_n)$ following (\ref{eq_SINR}), bit error rate as a function of SINR $\gamma_n$~\cite{1223549}, or probability of successful transmission (equivalently, effective channel capacity~\cite{9130689}).
$\overline{q}_n$ is the required minimum QoS level for link $n$, and $\overline{P}_n$ ($\underline{P}_n$) is the maximum (minimum) transmit power that can be adopted by SS $n$. Without loss of generality, we assume that $q_n(\mathbf{p}, \mathbf{u}_n, \mathbf{w}_n, \mathbf{u}_{-n}. \mathbf{w}_{-n})$ is monotonically increasing w.r.t. the local transmit power $P_n$ and decreasing w.r.t. the transmit power $P_i$ of any adversary link $i$ ($i\ne n$). Note that here we do not require any convexity property of $q(\cdot)$.

When central coordination is not available among the SS-DS links, as in a practical mmWave backhaul communication scenario, the original social optimization problem given in (\ref{eq_centralized_optimal}) can be re-formulated as a group of local optimization problems from a link-centric perspective. By assuming that each SS-DS link is self-interested in optimizing its own energy efficiency, we obtain the decentralized joint beamforming and power control problem from (\ref{eq_centralized_optimal}) as follows for SS-DS link $n=1,\ldots, N$:
\begin{subequations}\label{eq_local_optimal}
\begin{align}
  \tag{\ref{eq_local_optimal}}
    (P^*_n, \mathbf{u}^*_n, \mathbf{w}^*_n)= &
    \arg\min\limits_{P_n, \mathbf{u}_n, \mathbf{w}_n} c_n(P_n, P_{-n}) \\
    \label{eq_local_optimal_a}
    \textrm { s.t.}\quad & q_n(P_n, \mathbf{u}_n, \mathbf{w}_n, P_{-n}, \mathbf{u}_{-n}, \mathbf{w}_{-n}) \ge \overline{q}_n, \\
    \label{eq_local_optimal_b}
    & \underline{P}_n\le P_n\le \overline{P}_n,\\
    \label{eq_local_optimal_c}
    & \Vert \mathbf{w}_n \Vert = 1.
\end{align}
\end{subequations}
It is worth noting that the generalized QoS constraints in (\ref{eq_local_optimal_a}) and the nonlinear equality constraint in (\ref{eq_local_optimal_c}) render the local optimization problem a non-convex one (cf. (\ref{eq_central_optimal_a}) and (\ref{eq_central_optimal_c})). Then, proper relaxation-based methods should be sought for deriving the optimal solution of the local problem for each link as defined in (\ref{eq_local_optimal}). However, due to strategy coupling, link $n$ changing its local strategy $(P_n, \mathbf{u}_n, \mathbf{w}_n)$ will lead to the change in both the QoS and the cost functions of the other links $i\ne n$. Therefore, the decentralized problem of energy efficiency optimization in (\ref{eq_local_optimal}) can be described in the framework of non-cooperative games, where the strategy space of one link is determined by the joint action of the adversary links. Our focus will be developing a decentralized scheme of iterative power allocation and beamforming, which is able to approximate/converge to a fixed point (i.e., an NE) without the need of information exchange between links.

\subsection{Practical Case of the Generalized Problem in mmWave Infrastructure Networks}
\begin{figure}[t!]
  \centering
  \includegraphics[width=0.30\textwidth]{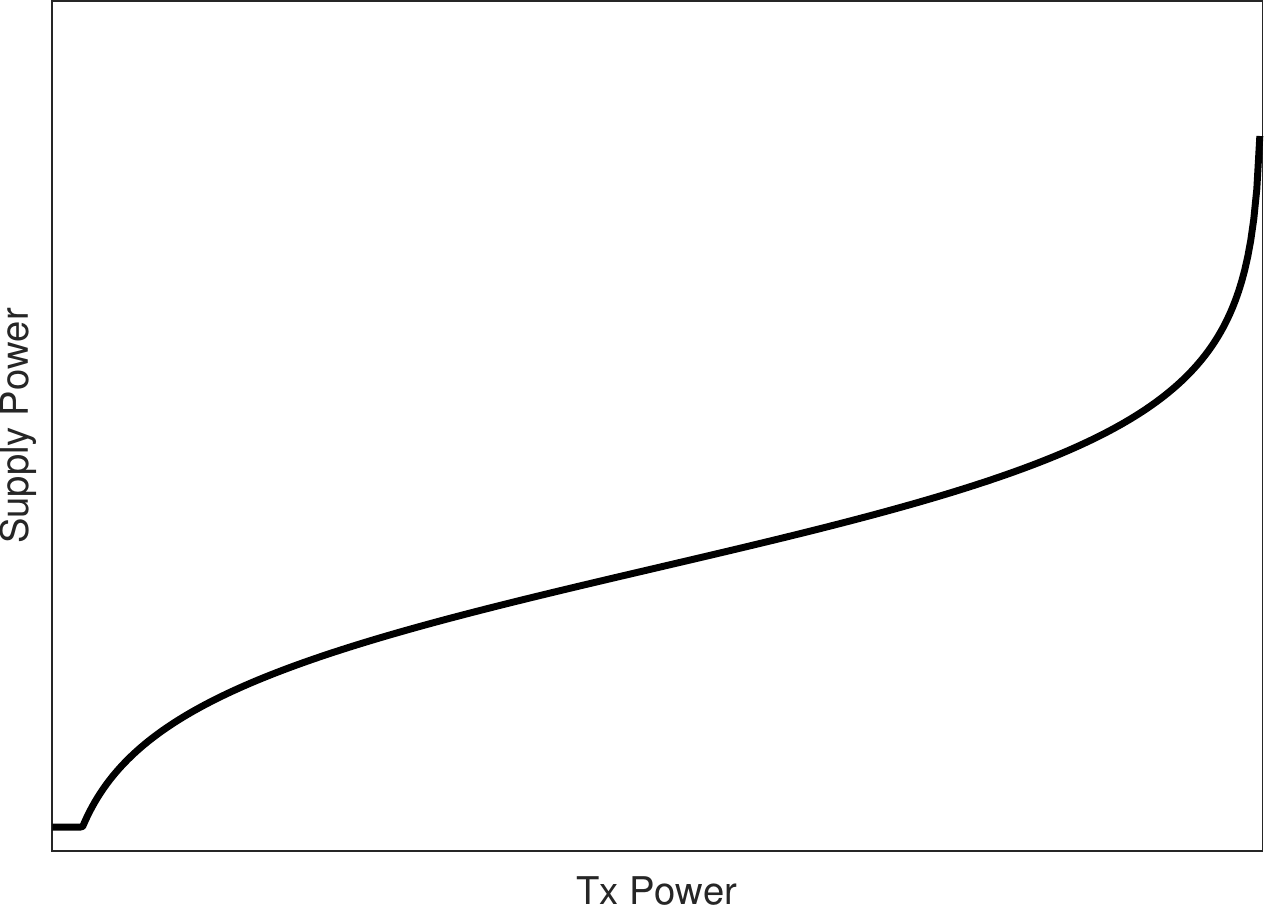}
  \caption{Illustration of the supply power as a function of the transmit power (cf.~\cite[Figures 6]{jensen2012lte}) in (\ref{eq_energy_efficiency}).}
  \label{figure_Illu_Supply_vs_Tx_Power}
\end{figure}
Based on the empirical studies in~\cite{6056587} and~\cite{jensen2012lte} (e.g.,~\cite[Figures 6]{jensen2012lte}), the supply power of an SS can be modeled in the form of an S-shaped function of the transmit power. More specifically, the supply power is mainly determined by the linear power consumption of the baseband and the nonlinear power consumption of the Power Amplifier (PA) due to input saturation at the RF chain (see Figure~\ref{figure_Illu_Supply_vs_Tx_Power}). Therefore, for the cost function $c_n(P_n, P_{-n})$ in the generalized local energy-efficiency problem as given in (\ref{eq_local_optimal}), we adopt the following energy efficiency model\footnote{While (\ref{eq_energy_efficiency}) is independent of the adversary power strategies $P_{-n}$, the generalized cost function $c_n(P_n, P_{-n})$ with power coupling can be conveniently used for modeling the adversary-dependent power consumption scenarios such as wireless energy harvesting.}:
\begin{equation}
  \label{eq_energy_efficiency}
  c_n(P_n) = \left\{
  \begin{array}{ll}
    \mu-\displaystyle\frac{1}{\alpha}\log({\overline{P}_n}/{P_n}-1), \textrm{ if } P_n > \underline{P}_n, \\
    \mu-\displaystyle\frac{1}{\alpha}\log({\overline{P}_n}/{\underline{P}_n}-1), \textrm{ if } P_n \le \underline{P}_n,
  \end{array}\right.
\end{equation}
where $\overline{P}_n$ is the upper bound for the achievable transmit power (due to PA saturation) of SS $n$, $\underline{P}_n$ is the minimum effective transmit power that can be achieved by the circuit, $\mu$ is the shifter parameter to reflect the impact of the baseband power, and $\alpha$ is the slope parameter to reflect the PA characteristics. The piece-wise expression of $c_n(P_n)$ is adopted to capture the power level required for maintaining the circuit.
We consider that the QoS over link $n$ is measured as a non-decreasing function of the local SINR (e.g., BER or throughput), namely, $q_n(P_n, \mathbf{u}_n, \mathbf{w}_n, P_{-n}, \mathbf{u}_{-n}, \mathbf{w}_{-n})=q_n(\gamma_n)$ based on (\ref{eq_SINR}). In this case, the inequality constraint in (\ref{eq_local_optimal_a}) can be rewritten into a linear form w.r.t. $P_i$ ($\forall i=1,\ldots, N$) as follows:
\begin{equation}
  \label{eq_linear_inequality}
    P_n\Vert \mathbf{u}^{\textrm{H}}_n\mathbf{H}_{n,n}\mathbf{w}_n\Vert^2 -\overline{\gamma}_n\bigg(\Vert\mathbf{u}^{\textrm{H}}_n\eta_n\Vert^2 + \sum\limits_{i\ne n} P_i\Vert \mathbf{u}^{\textrm{H}}_n\mathbf{H}_{i,n}\mathbf{w}_i\Vert^2\bigg)\ge 0,
\end{equation}
where $\overline{\gamma}_n$ indicates the threshold value of the minimum SINR that corresponds to $\overline{q}_n$ in (\ref{eq_local_optimal_a}). Consequently, the generalized local optimization problem in (\ref{eq_local_optimal}) now becomes
\begin{subequations}\label{eq_local_optimal_instance}
\begin{align}
  \tag{\ref{eq_local_optimal_instance}}
    (P^*_n, \mathbf{u}^*_n, \mathbf{w}^*_n) & = \arg\min\limits_{P_n, \mathbf{u}_n, \mathbf{w}_n} c_n(P_n) \\
    \textrm {s.t.}\;\;\; & P_n\Vert \mathbf{u}^{\textrm{H}}_n\mathbf{H}_{n,n}\mathbf{w}_n\Vert^2 - \nonumber \\
    \label{eq_local_optimal_a_instance}
    & \overline{\gamma}_n\Big(\Vert\mathbf{u}^{\textrm{H}}_n\eta_n\Vert^2 + \sum\limits_{i\ne n} P_i\Vert  \mathbf{u}^{\textrm{H}}_n\mathbf{H}_{i,n}\mathbf{w}_i\Vert^2\Big)\!\ge \!0, \\
    \label{eq_local_optimal_b_instance}
    & \underline{P}_n\le P_n\le \overline{P}_n, \\
    \label{eq_local_optimal_c_instance}
    & \Vert \mathbf{w}_n \Vert = 1.
\end{align}
\end{subequations}

Note that with the objective function in (\ref{eq_energy_efficiency}) and the equality constraint (\ref{eq_local_optimal_c_instance}) (see also (\ref{eq_local_optimal_c})), the local power allocation problem in (\ref{eq_local_optimal_instance}) is a non-convex one. The joint power-beamformer optimization problem in (\ref{eq_local_optimal_instance}) has been a relatively well-studied problem, and this allows us to compare to other techniques in the literature (see also our discussion in Section~\ref{Sec:Simulation}). To demonstrate the expressiveness of our  framework of decentralized solution based on the formulation of generalized games, we provide another example of constructing the local optimization problem with non-convex QoS constraints in Appendix~\ref{app_non_convex_example}.

\section{A Decentralized Approach to the Formulated Problem}\label{Sec:Dec_approach}
In this section, we focus on the solution to the decentralized power-efficiency problem given in (\ref{eq_local_optimal}). We note that the objective function $c_n(P_n, P_{-n})$ in (\ref{eq_local_optimal}) is independent of the beamforming vectors. Thus, we are able to divide the optimization process for joint power allocation and beamforming into two stages. In the first stage, with the temporarily fixed Tx/Rx filters for all the links, we tackle the original problem given in (\ref{eq_local_optimal}) as a sub-problem of decentralized search for the locally optimal power allocation. Based on the formulation of a non-cooperative, generalized Nash game, we propose a purely decentralized strategy-search scheme, which theoretically guarantees the convergence to the Generalized NE (GNE) of the power strategies (furthermore, which is socially optimal under certain condition). In the second stage, for each SS-DS link we introduce the MSE criterion for Rx/Tx beamformer design. We propose a joint power allocation and beamforming scheme in the form of iterative adaptation of the Rx/Tx beamformers, using the power allocation strategies obtained in the GNE game of the first stage as the input. Additionally, we also investigate a series of Tx-beamforming schemes based on different levels of information exchange between links, and provide a theoretical comparison of their impact on the convergence of the proposed two-stage strategy-searching algorithm.

\subsection{Game-Theoretic Solution for the Sub-problem of Power Allocation}
We first consider that the fixed Tx/Rx-beamforming vectors $\mathbf{w}_n$, $\mathbf{u}_n$ are adopted by each SS-DS link $n$ in the ad-hoc network. Then, the generalized local joint beamforming and power allocation problem as defined in (\ref{eq_local_optimal}) is reduced into the following form for $n=1,\ldots, N$:
\begin{subequations}\label{eq_local_optimal_fixed_bfv}
\begin{align}
  \tag{\ref{eq_local_optimal_fixed_bfv}}
    P^*_n=\arg\max\limits_{ P_n} &\; \left\{u_n(P_n, P_{-n})=-c_n(P_n, P_{-n})\right\} \\
    \label{eq_local_optimal_fixed_bfv_a}
    \qquad\textrm {s.t.}\;\; & \; q(P_n, P_{-n}) \ge \overline{q}_n, \\
    \label{eq_local_optimal_fixed_bfv_b}
    &\; \underline{P}_n\le P_n\le \overline{P}_n.
\end{align}
\end{subequations}

Since the feasible strategy set of each player depends on the strategy of the adversary players (see (\ref{eq_local_optimal_fixed_bfv_a})), we are ready to model the interaction among the links with the local goal of (\ref{eq_local_optimal_fixed_bfv}) as a generalized Nash game of $N$ players~\cite{facchinei2010generalized}, where the set of $N$ links in the network, $\mathcal{N}\!=\!\{1,\ldots, N\}$, are considered as the players of the game. For player $n$, the feasible strategy set of its power allocation $P_n$ is determined by the joint strategy adopted by the adversary players $P_{-n}$:
\begin{equation}
  \label{eq_feasible_strategy}
  P_n\in\mathcal{P}_n(P_{-n})=\{P_n: \underline{P}_n\le P_n\le \overline{P}_n, q_n(P_n, P_{-n})\ge\overline{q}_n \}.
\end{equation}
Thereby, we can denote the strategy set of the game as a point-to-set mapping $\mathcal{P}(\mathbf{p})=\prod_{n=1}^{N}\mathcal{P}_n(P_{-n})$. The formulation of the generalized Nash game\footnote{$\mathcal{G}$ is a generalized game, since by (\ref{eq_feasible_strategy}) the strategy spaces of each player strongly depends on the strategies of the other layers and is not fixed as in standard normal-form games.} is completed as follows:
\begin{Definition}
  \label{def_GNE}
  The generalized Nash game for power allocation is a 3-tuple $\mathcal{G}=\left\langle \mathcal{N}, \mathcal{P}(\mathbf{p}), U\right\rangle$, where $\mathcal{N}\!=\!\{1,\ldots, N\}$ is the set of players, $\mathcal{P}(\mathbf{p})=\prod_{n=1}^{N}\mathcal{P}_n(P_{-n})$ is the feasible set of joint power profiles and $U=(u_1,\ldots, u_N)$ is the vector of utility functions as defined in (\ref{eq_local_optimal_fixed_bfv}).

  A GNE in the game $\mathcal{G}$ finds the simultaneous power allocation such that $\mathbf{p}^*\in\mathcal{P}(\mathbf{p}^*)$ and $\forall n=1,\ldots, N$,
  \begin{equation}
    \label{eq_tuple_game}
    P_n^*=\arg\max_{P_n} u_n(P_n, P_{-n}) \textrm{ s.t. } P_n\in\mathcal{P}_n(P_{-n}).
  \end{equation}
\end{Definition}

According to the theory of the GNE problem~\cite{facchinei2010generalized}, to ensure the existence of such a GNE, we typically need the local feasible strategy sets $\mathcal{P}_n(P_{-n})$ to be non-empty, closed and convex, and  the utility function $u_n(P_n, P_{-n})$ to be quasi-concave w.r.t. $P_n$ on $\mathcal{P}_n(P_{-n})$ for all the players. However, this condition may not necessarily be satisfied\footnote{For the specific local problem using linear-interference-based QoS measurement as given in (\ref{eq_local_optimal_instance}), we know that $\forall n\in\mathcal{N}$, if its feasible strategy space $\mathcal{P}_n(P_{-n})$ is non-empty, $\mathcal{P}_n(P_{-n})$ is closed and convex. However, we know from~\cite{5710988} that the feasible SINR region is in general non-convex for a generalized non-linear-interference-based QoS function.} for the category of problems we are considering with the generalized model in (\ref{eq_local_optimal_fixed_bfv}). As a result, the standard KKT condition-based methods~\cite[Chapter 10]{facchinei2007finite} for GNE solution may not apply, and we need to find an alternative approach for the analysis and search of the GNE points. In this paper, we are interested in a special family of generalized Nash games that satisfy the following properties:
\begin{Definition}
  \label{def_monotonic_properties}
  A generalized Nash game $\mathcal{G}=\left\langle \mathcal{N}, \mathcal{P}(\mathbf{p}), U\right\rangle$ is monotonic if the following properties hold:
  \begin{itemize}
    \item[(a)] $\forall n\in\mathcal{N}$, $u_n(P_n, P_{-n})$ is continuous and monotonically decreasing in $P_n$.
    \item[(b)] $\forall n\in\mathcal{N}$, the local feasibility set $\mathcal{P}_n(P_{-n})=\{P_n: \underline{P}_n\le P_n\le \overline{P}_n, q_n(P_n, P_{-n})\ge\overline{q}_n \}$ satisfies the condition that $q_n(P_n, P_{-n})$ is continuous, monotonically increasing in $P_n$ and monotonically decreasing in each $P_{i}$ ($\forall i\ne n$).
  \end{itemize}
\end{Definition}
Note that since $\forall n\in\mathcal{N}$, $P_n$ is bounded and $q_n(P_n,P_{-n})$ is continuous, $\mathcal{P}(\mathbf{p})$ is compact. Meanwhile, the ``monotonicity porperty'' in Definition~\ref{def_monotonic_properties} differs from the property of ``monotone mapping over convex and compact set'' required by equilibrium searching in typical convex games with coupled feasibility constraints (e.g., the monotone pseudo-gradient mapping in~\cite{9525284}). Comparatively, our considered games do not need to be convex to allow for the development of equilibrium searching methods.

\subsection{Analysis and Distributed Search of Generalized Nash Equilibria}
\label{subsec:analysis_NE}
We first provide the existence condition for the considered generalized power allocation game $\mathcal{G}$ in Theorem~\ref{thm_existence}. The development of Theorem~\ref{thm_existence} is constructive and relies on the monotonicity property of the considered game (see Definition~\ref{def_monotonic_properties}). Then, with Theorem~\ref{thm_convergence}, we propose an algorithm which converges to the GNE when a feasible solution exists or identifies that the problem is infeasible.
\begin{Theorem}[Existence]
  \label{thm_existence}
  Assume that $\mathcal{G}=\left\langle \mathcal{N}, \mathcal{P}(\mathbf{p}), U\right\rangle$ is monotonic, namely, satisfies the conditions given in Definition~\ref{def_monotonic_properties}.
  Then, if a feasible strategy profile $\hat{\mathbf{p}}$ exists, i.e., $\hat{P}_n\in\mathcal{P}(P_{-n})$, there exists a GNE point satisfying (\ref{eq_tuple_game}) for all $n\in\mathcal{N}$.
\end{Theorem}
\begin{proof}
  The proof to Theorem~\ref{thm_existence} is composed of two parts. We first show that there exists a convergent sequence of feasible strategies starting from the feasible strategy profile $\hat{\mathbf{p}}$ with simultaneous strategy updates with local best response. Then, we show that the limit of the convergent sequence is a GNE.

  Assume that a strategy updating sequence $\{\mathbf{p}(t)\}^{\infty}_{t=0}$ is initialized with $\mathbf{p}(0)=\hat{\mathbf{p}}$. By monotonicity of $u_n(P_n,P_{-n})$ and $q_n(P_n,P_{-n})$, we know that at time $t$, a feasible local best response for player $n$, $P_n(t+1)$, exists and either satisfies the equality condition $q_n(P_n(t+1), P_{-n}(t))=\overline{q}_n$, or $P_n=\underline{P}_n$. In the latter case, (using the monotonicity assumptions) any further reduction of power by other players will not change the feasibility with minimal power for this specific player. Hence we now assume that the equality condition holds for the rest of the proof. By construction, the sequence $\{\mathbf{p}(t)\}^{\infty}_{t=0}$ is updated simultaneously through local best response. Equivalently, in $\{\mathbf{p}(t)\}^{\infty}_{t=0}$, the strategy at $t+1$, $\mathbf{p}(t+1)$, is updated for each element $n\in\mathcal{N}$ following
  \begin{equation}
    \label{eq_br_updating_proof_existence}
    P_n(t+1)=\arg\min_{P}\{P: q_n(P, P_{-n}(t))\ge\overline{q}_n, \underline{P}_n\le P\le \overline{P}_n\}.
  \end{equation}

  Since $q_n(P_n,P_{-n})$ is monotonically increasing in $P_n$ and monotonically decreasing in $P_i$ for all $i\ne n$, we know from the updating rule in (\ref{eq_br_updating_proof_existence}) that $P_n(t+1)\le P_n(t)$, $\forall n\in\mathcal{N}$. Then, by construction of $\{\mathbf{p}(t)\}^{\infty}_{t=0}$ with (\ref{eq_br_updating_proof_existence}), $\mathbf{p}(t)$ is element-wise monotonically decreasing and thus feasible for all $t$, namely,
  \begin{equation}
    \label{eq_sequence_feasibility}
    \forall t, \forall n\in\mathcal{N}: q_n(P_n(t), P_{-n}(t))\ge\overline{q}_n.
  \end{equation}
  Furthermore, by compactness of $\mathcal{P}(\mathbf{p})$, we have
  \begin{equation}
    \label{eq_sequence_limit_existence}
    \lim_{t\rightarrow\infty}\mathbf{p}(t)=\mathbf{p}^{\infty}.
  \end{equation}
  By continuity of $q_n(P_n, P_{-n})$, $\mathbf{p}^{\infty}$ is also feasible, and we have $\lim\limits_{t\rightarrow\infty}q_n(P_n(t),P_{-n}(t))=q_n(P_n^{\infty},P_{-n}^{\infty})$.

  To show that ${\mathbf{p}}^{\infty}$ is a GNE, we assume that there exists at least one link $n$ such that $q_n(P_n^{\infty},P^{\infty}_{-n})>\overline{q}_n$ and $P_n^{\infty}\ne\underline{P}_n$. By monotonicity of $\mathbf{p}(t)$ over all $t$,
  \begin{equation}
    \label{eq_contradiction_existence_1}
    q_n(P_n^{\infty},P^{\infty}_{-n})>q_n(P_n^{\infty}, P_{-n}(t)).
  \end{equation}
  By continuity of $q_n(\cdot)$, a sufficiently large $t_0$ exists such that
  \begin{equation}
    \label{eq_contradiction_existence}
    q_n(P_n^{\infty}, P_{-n}(t_0))>\overline{q}_n.
  \end{equation}
  This is a contradiction, since this implies by construction of the sequence with (\ref{eq_br_updating_proof_existence}) that $P_n^{\infty}>P_n(t_0+1)$. Therefore, by monotonicity of $u_n(P_n,P_{-n})$, simultaneous best response is reached at ${\mathbf{p}^{\infty}}$, and thus ${\mathbf{p}^{\infty}}$ is a GNE. This completes the proof to Theorem~\ref{thm_existence}.
\end{proof}

\begin{Algorithm}[!t]
  \centering
\begin{tikzpicture}[node distance=1.5cm, align=center]
  \tikzstyle{startstop} = [rectangle, rounded corners, minimum width=1cm, minimum height=0.5cm,text centered, draw=black, fill=gray!30]
  \tikzstyle{io} = [trapezium, trapezium left angle=70, trapezium right angle=110, minimum width=1cm, minimum height=0.5cm, text centered, draw=black, fill=gray!30]
  \tikzstyle{process} = [rectangle, minimum width=1cm, minimum height=0.5cm, text centered, draw=black, fill=gray!30]
  \tikzstyle{decision} = [diamond, minimum width=0.5cm, minimum height=0.5cm, text centered, draw=black, fill=gray!30]

  \tikzstyle{arrow} = [thick,->,>=stealth]
  %\node (start) [startstop] {\scriptsize Start};
  \node (initialization) [startstop, text width = 2.5cm,] {\tiny\linespread{0.8} SS $n$ sets $\mathbf{u}_n$, $\mathbf{w}_n$, $P_n\!=\!\underline{P}_n$ and $\delta>0$\par};
  \node (pro_transmit) [process, text width = 2.5cm, below of= initialization, yshift=0.4cm] {\tiny Transmit pilot with $P_n$};
  \node (pro_waitack) [process, text width = 2.5cm, below of= pro_transmit, yshift=0.4cm] {\tiny Receive $\textrm{Ack}_n\in\{0,1\}$ from DS $n$\par};

  \node (dec_ack) [decision, aspect=2.5, below of= pro_waitack, yshift=0.3cm] {\tiny $\textrm{Ack}_n = 0$\par};
  \node (dec_maxP) [decision, aspect=2.5, below of= dec_ack, yshift=0.2cm] {\tiny $P_n=\overline{P}_n$\par};

  \node (pro_update) [process, text width = 2.5cm, right of= dec_maxP, xshift=1.5cm] {\tiny Compute $P'_n$ according to (\ref{eq:def_Pt})\par};
  \node (dec_epsilonNE) [decision, aspect=2.5, right of= dec_ack, xshift=1.5cm] {\tiny $\vert P'_n\!-\!P_n\vert\!\le\!\delta$\par};
  \node (pro_waitpilot) [process, text width = 2.0cm, right of= dec_epsilonNE, xshift=1.5cm] {\tiny Listen to SS-Notification Channel (SS-NC)\par};

  \node (pro_occupyssnc) [process, text width = 2.5cm, right of= pro_waitack, xshift=1.5cm] {\tiny Transmit over SS-NC (blocking)\par};

  \node (dec_ssncbusy) [decision, aspect=2.5, right of=pro_occupyssnc, xshift=1.5cm] {\tiny Is SS-NC busy?\par};
  \node (pro_setnew) [process, text width = 2.5cm, right of= pro_transmit, xshift=1.5cm] {\tiny Update $P_n = P'_n$\par};

  \node (output) [startstop, text width = 2.5cm, yshift=0.1cm, below of=dec_maxP, yshift=0.2cm] {\tiny Terminal: SS $n$ output $P_n$\par};

  \draw [arrow] (initialization) -- (pro_transmit);
  \draw [arrow] (pro_transmit) -- (pro_waitack);
  \draw [arrow] (pro_waitack) -- (dec_ack);

  \draw [arrow] (dec_ack) -- node [anchor=east]{\tiny Yes} (dec_maxP);
  \draw [arrow] (dec_maxP) -- node [anchor=east]{\tiny Yes} (output);

  \draw [arrow] (dec_ack) ($(dec_ack.east)$) -- node [anchor=north]{\tiny No} (pro_update);
  \draw [arrow] (dec_maxP) ($(dec_maxP.east)$) --  node [anchor=north]{\tiny No} (pro_update);

  \draw [arrow] (pro_update) -- (dec_epsilonNE);
  \draw [arrow] (dec_epsilonNE) -- node [anchor=east]{\tiny No} (pro_occupyssnc);
  \draw [arrow] (dec_epsilonNE) -- node [anchor=south]{\tiny Yes} (pro_waitpilot);

  \draw [arrow] (pro_waitpilot) -- (dec_ssncbusy);
  \draw [arrow] (pro_occupyssnc) -- (dec_ssncbusy);

  \draw [arrow] (dec_ssncbusy) |- node [anchor=west]{\tiny Yes} (pro_setnew);
  \draw [arrow] (pro_setnew) -- (pro_transmit);

  \draw [arrow] (dec_ssncbusy) ($(dec_ssncbusy.east)$) |- node [anchor=west]{\tiny No} (output);

%  \draw [arrow] (dec4) -- node [anchor=east]{\tiny No} (pro_update);
%  \draw [arrow] (dec4)  |- node [anchor=north]{\tiny Yes} (out1);

%  \draw [arrow] (dec3) -- node [anchor=east]{\tiny No} (pro_setnew);
%  \draw [arrow] (pro_setnew) -- (pro_transmit);
%  \draw [arrow] (dec3) ($(dec3.east)$) -| node [anchor=west]{\tiny Yes} (out2);
\end{tikzpicture}
\caption{Protocol sketch of asynchronous strategy update for link $n$ with fixed $\mathbf{u}_n$ and $\mathbf{w}_n$. $\textrm{Ack}_n=0$ is sent by DS $n$ upon unsatisfactory QoS levels. SS $n$ notifies the other links of the completion of iteration by evacuating from the SS-Notification channel.}
  \label{alg_jacobian}
\end{Algorithm}

The constructed monotonic strategy sequence in the proof of Theorem~\ref{thm_existence} indicates that each time a player reduces its power the others benefit, so they can further reduce the power. Hence no matter what the order of strategy updates is, the process will converge. Inspired by the finding obtained in Theorem~\ref{thm_existence}, we aim to develop a distributed algorithm that is able to form a similar convergent strategy sequence without requiring identifying a feasible joint strategy for initialization. Thus, we design an asynchronous scheme for strategy updating in the non-cooperative game $\mathcal{G}$ in Algorithm~\ref{alg_jacobian}. In Theorem~\ref{thm_convergence}, we further show that the solution feasibility of any instance of the general monotonic power allocation game can be identified following Algorithm~\ref{alg_jacobian}. Furthermore, Algorithm~\ref{alg_jacobian} is guaranteed to converge in finite time whenever the set of GNE in the game $\mathcal{G}$ is non-empty and utilities and QoS functions are Lipschitz.

\begin{Theorem}[Convergence]
  \label{thm_convergence}
    Assume that $\mathcal{G}=\left\langle \mathcal{N}, \mathcal{P}(\mathbf{p}), U\right\rangle$ is monotonic as defined in Definition~\ref{def_monotonic_properties}. Then, by initializing ${P}_n(0)={\underline{P}_n}$ for each player $n\in\mathcal{N}$ and for an accuracy parameter $\epsilon>0$, there exists a $\delta>0$ such that Algorithm~\ref{alg_jacobian} terminates in finitely many iterations with the condition $\forall n\in\mathcal{N}:\Vert P_n(t+1)-P_n(t)\Vert\le \delta$ by either identifying that no
    feasible allocation exists or converging to an $\varepsilon$-GNE\footnote{At an $\varepsilon$-GNE $(P^*_n,P^*_{-n})$, no player (e.g., $n$) can increase its utility by more than $\varepsilon$ with any feasible strategy $P_n$, namely $u_n(P_n, P^*_{-n})- u_n(P^*_n, P^*_{-n})\le\varepsilon$, $\forall n\in\mathcal{N}$ and $\forall P_n\in\mathcal{P}(P^*_{-n})$.}.

    Moreover, If all the functions, $q_n(\cdot), c_n(\cdot)$ are $L$ Lipschitz with respect to $\|\cdot\|_{\infty}$, then we can choose $\delta=\frac{\varepsilon}{L}$ and the algorithm converges within $
    {\delta^{-1}NL\max\limits_{n\in\mathcal{N}}\left(\overline{P}_n-{\underline P}_n\right)}$ iterations.
\end{Theorem}
\begin{proof}
  Consider that the strategy sequence $\{\mathbf{p}(t)\}^{T}_{t=0}$ is initialized as $\mathbf{p}(0)=[\underline{P}_1,\ldots,\underline{P}_N]^{\textrm{T}}$, and then updated following the rule as defined in (\ref{eq:def_Pt}) (cf. (\ref{eq_br_updating_proof_existence})):
  \begin{align}
    & P_n(t+1)= \nonumber \\
    & \arg\min_{P}\{P: q_n(P, P_{-n}(t))\ge\overline{q}_n+\gre, \underline{P}_n(t)\le P\le \overline{P}_n\}.
    \label{eq:def_Pt}
  \end{align}
  Let $\vp(t)=\left[ P_1(t),\ldots,P_N(t)\right]$, then, by monotoniticy of $q_n(\vp)$, at each stage, each player increases its power to the minimal level conditioned on the current strategies of the other players and the QoS constraint condition in (\ref{eq:def_Pt}). Since all $q_n(\cdot),u_n(\cdot)$ are continuous, the set of feasible points is either empty (if the problem is infeasible) or compact. Assume that the problem is feasible. Then, for all $n\in\mathcal{N}$, $q_n(\cdot)$ and $u_n(\cdot)$ are uniformly continuous.
    Let $\gre>0$ be given. We can always choose a $\gd>0$ such that for all $n$ if $\|\vp-\vp'\|_{\infty}<\gd$ then, for all $n\in\mathcal{N}$
    \begin{align}
        \Vert q_n(\vp)-q_n(\vp')\Vert <\gre \textrm{ and }
        \Vert u_n(\vp)-u_n(\vp')\Vert <\gre.
        \label{eq_termination_condition}
    \end{align}
    Let $\gD P=\max_n\left(\overline{P}_n-\underline{P}_n\right)$. By the stopping condition $\|\vp-\vp'\|_{\infty}<\gd$, the algorithm must terminate after at most $\gd^{-1}N\gD P$ iterations, since if the algorithm did not terminate, at least one player would have increased its power by more than $\gd$ and after at most $\gd^{-1}\gD P$ iterations it becomes infeasible.

   If for some time $t+1$, no player increases its power by more than $\gd$, we know by the uniform continuity that for all $n\in\mathcal{N}$
   \begin{align}
   \|q_n(\vp(t+1))-q_n(P_{n}(t+1),P_{-n}(t))\|_{\infty}<\gre.
   \end{align}
   By the definition of the updating rule for $P_{n+1}(t)$ in (\ref{eq:def_Pt}), we know that either
   \begin{align}
       q_n(P_{n}(t+1),P_{-n}(t))\ge \overline{q}_n+\gre
   \end{align}
   and the player increases its power by less than $\gd$, or
      \begin{align}
       q_n(P_{n}(t),P_{-n}(t))\ge \overline{q}_n+\gre,
   \end{align}
   if it is satisfied with the previous power.
Therefore, $q_n(\vp(t+1))\ge \overline{q}_n$. By the selection of $\gd$ and monotonicity of $u_n(P)$, we also know that $u_n(P_n(t+1))<u_n(P_n(t))+\gre.$ Therefore, $\vp(t+1)$ is an $\gre$-Nash.

Moreover, if $\forall n\in\mathcal{N}$, $q_n(\cdot)$ and $u_n(\cdot)$ are $L$ Lipschitz, from the above discussion and by the definition of Lipschitz continuity we have
    \begin{align}
    \label{eq_uniform_continuity_all_Lipschitz}
    \forall n\in\mathcal{N}:\; & \Vert q_n(\vp(t+1)) - {q}_n(P_n(t+1), P_{-n}(t))\Vert_{\infty} \nonumber  \\
    & <L\Vert \mathbf{p}(t+1)-\mathbf{p}(t)\Vert_{\infty},
  \end{align}
  Then, by setting the stopping condition $\delta<\varepsilon/L$, from (\ref{eq_uniform_continuity_all_Lipschitz}) we know that the $\varepsilon$-GNE can be reached within $\delta^{-1}{NL\max\limits_{n\in\mathcal{N}}(\overline{P}_n-\underline{P}_n)}$ iterations.
%
%  by continuity of $\prod_{n\in\mathcal{N}}[\underline{P}_n, \overline{P}_n]$, we can always find $t<\infty$, such that $\forall t'\ge t$, $\Vert \mathbf{p}(t') - \mathbf{p}^{\infty}\Vert_{\infty}<\varepsilon$. Then, by definition of $\varepsilon$-GNE and continuity of $u_n(P_n,P_{-n})$, $\mathbf{p}(t)$ is an $\varepsilon$-GNE. Therefore, the proof is completed.
\end{proof}

Now, we consider the instantiation of the generalized joint power allocation and beamforming problem presented in (\ref{eq_local_optimal_instance}). By exploiting the special structure of the link-cost function, we can show in Corollary~\ref{thm_optimal} that the fixed-point strategies obtained with Algorithm~\ref{alg_jacobian} align with the socially optimal strategies in the sub-problem of the power allocation game as derived from (\ref{eq_local_optimal_instance}) by fixing the beamforming vectors.
\begin{Corollary}
  \label{thm_optimal}
  Assume that $\mathcal{G}=\left\langle \mathcal{N}, \mathcal{P}(\mathbf{p}), U\right\rangle$ is a monotonic generalized Nash game, where $u(P_n, P_{-n})$ is independent of the adversary power strategy $P_{-n}$ (as in (\ref{eq_local_optimal_instance})).
  Then, if a GNE $\mathbf{p}^*$ is reached by Algorithm~\ref{alg_jacobian} from $\mathbf{p}(t=0)=[\underline{P}_{n:n\in\mathcal{N}}]^{\textrm{T}}$, $\mathbf{p}^*$ is socially optimal.
\end{Corollary}
\begin{proof}
  Assume $\mathbf{p}^*$ is a GNE such that $\mathbf{p}^*$ is a socially non-optimal strategy. Assume that $\tilde{\mathbf{p}}$ is a feasible strategy with
  \begin{equation}
    \label{eqn_inequality_optimal_solution}
    \sum_{n=1}^N u_n(P^*_n,P^*_{-n}) < \sum_{n=1}^N u_n(\tilde{P}_n, \tilde{P}_{-n}).
  \end{equation}
  Then, by the decreasing monotonicity of $u_n({P}_n, {P}_{-n})=u_n({P}_n)$, there exists at least one $n$ such that $u_n(P^*_n)<u_n(\tilde{P}_n)$ and thus $\tilde{P}_n<P^*_n$. By the monotonicity property of $q_n(P_n, P_{-n})$ in $P_n$ we can observe that $(\tilde{P}_n, P^*_{-n})$ is also feasible, i.e., $\forall i\in\mathcal{N}:q_i(P_n, P_{-n})\ge\overline{q}_i$, and $u_n(P^*_n,P^*_{-n})<u_n(\tilde{P}_n, P^*_{-n})$. This contradicts with the assumption that $\mathbf{p}^*$ is a GNE. Therefore, $\tilde{\mathbf{p}}$ does not exist, which complete the proof to Corollary~\ref{thm_optimal}.
\end{proof}
Then, if the cost function of each link satisfies the condition as given by Definition~\ref{def_monotonic_properties} and Corollary~\ref{thm_optimal}, the NE of the formulated power allocation game $\mathcal{G}$, as obtained following Algorithm~\ref{alg_jacobian}, conveniently provides the socially optimal solution to the problem as given in (\ref{eq_centralized_optimal}).

\begin{Remark}
  It is worth noting that Algorithm~\ref{alg_jacobian} works properly regardless of the synchronization levels of user actions. Note that the convergence property of conventional potential games (e.g., based on best response), cannot be directly applied to a non-specific generalized NE problem.
  %It is known that even if game $\mathcal{G}$, is a generalized ordinal potential game, Algorithm~\ref{alg_jacobian} may not lead to a GNE starting from an arbitrary feasible point~\cite{facchinei2011decomposition} with a generic constraint. The convergence property given by Corollary~\ref{thm_optimal} relies on the properties of the game as specified in conditions (a) and (b) of Definition~\ref{def_monotonic_properties}. Otherwise, the strict concavity of $u_n(P_n, P_{-n})$ over a convex, closed and compact point-to-set mapping $\mathcal{P}{(\mathbf{p})}$ must be satisfied by the game.

  Theorems~\ref{thm_existence} and~\ref{thm_convergence} are still valid when each player's utility also depends on $P_{-n}$, as long as $u_n(\mathbf{p})$ is decreasing in $P_n$. When $u_n(\mathbf{p})$ depends on $P_{-n}$ in an adversarial manner, i.e, it is monotonically decreasing with each coordinate of $P_{-n}$ as well, Corollary~\ref{thm_optimal} still holds such that a GNE yields a socially optimal solution, since the goals of the players are aligned. Each player minimizing its power is also increasing the utility of the other players. However, in more general cases, a GNE is no longer socially optimal.
\end{Remark}

\subsection{Joint Power Allocation and Beamforming}
\label{subsec:two_stage_and_convergence}
Now, we consider that the power allocation strategies of all the links are updated using Algorithm~\ref{alg_jacobian} in a one-shot GNE game. Then, to ensure that the receiving filter of DS $n$, $\mathbf{u}_n$ ($n\in\mathcal{N}$), satisfies the constraint condition on QoS in (\ref{eq_local_optimal_a}) with fixed Tx-beamformers $\mathbf{w}_n$ ($n\in\mathcal{N}$), we introduce the local MSE-based criterion for designing $\mathbf{u}_n$ as follows:
\begin{equation}
  \label{eq_mse}
  \mathbf{u}^*_n =\arg\min_{\mathbf{u}_n}\left\{\textrm{MSE}_n=E\left\{\Vert \hat{s}_n - {s}_n\Vert^2\right\}\right\},
\end{equation}
where $\hat{s}_n$ is obtained as $\hat{s}_n =\mathbf{u}^{\textrm{H}}_n\mathbf{y}_n$ following (\ref{eq_received_signal}).
When ${s}_n$ is an independent zero-mean complex Gaussian signal of unit variance $\forall n\in\mathcal{N}$, the objective function in (\ref{eq_mse}) can be expanded as follows:
  \begin{align}
    & \textrm{MSE}_n(P_n, \mathbf{u}_n; P_{-n})=\tr\left(E\left\{\Vert \hat{{s}}_n - {s}_n\Vert^2\right\}\right) \nonumber \\
    & =E\Bigg\{\tr\bigg[\bigg(\mathbf{u}^{\textrm{H}}_n\sum\limits_{i= 1}^N\sqrt{P_i}\mathbf{H}_{i,n}\mathbf{w}_i{s}_i -{s}_n + \mathbf{u}^{\textrm{H}}_n\boldsymbol{\eta}_n\bigg) \nonumber \\
    & \bigg(\mathbf{u}^{\textrm{H}}_n\sum\limits_{i= 1}^N\sqrt{P_i}\mathbf{H}_{i,n}\mathbf{w}_i{s}_i -{s}_n + \mathbf{u}^{\textrm{H}}_n\boldsymbol{\eta}_n\bigg)^{\textrm{H}}
    \bigg]
    \Bigg\}  \nonumber \\
    & =\tr\bigg(\mathbf{u}^{\textrm{H}}_n\left(\sum\limits_{i=1}^{N}P_i\mathbf{H}_{i,n}\mathbf{w}_i\mathbf{w}^{\textrm{H}}_i\mathbf{H}^{\textrm{H}}_{i,n}\right)\mathbf{u}_n-\sqrt{P_n}\mathbf{w}^{\textrm{H}}_n\mathbf{H}^{\textrm{H}}_{n,n}\mathbf{u}_n \nonumber \\
    & -\sqrt{P_n}\mathbf{u}^{\textrm{H}}_n\mathbf{H}_{n,n}\mathbf{w}_n+1+\sigma^2_n\mathbf{u}^{\textrm{H}}_n\mathbf{u}_n\bigg).
    \label{eq_mse_expand}
  \end{align}
Then, from the perspective of SS-DS link $n$, to minimize $\textrm{MSE}_n$ in (\ref{eq_mse_expand}), we need $\partial \textrm{MSE}_n/\partial \mathbf{u}_n=0$. This leads to
\begin{equation}
  \label{eq_mmse_filter}
  2\mathbf{u}^{\textrm{H}}_n\left(\sum\limits_{i=1}^{N}P_i\mathbf{H}_{i,n}\mathbf{w}_i\mathbf{w}^{\textrm{H}}_i\mathbf{H}^{\textrm{H}}_{i,n}\right)
  -2\sqrt{P_n}\mathbf{w}^{\textrm{H}}_n\mathbf{H}^{\textrm{H}}_{n,n}+2\sigma_n^2\mathbf{u}^{\textrm{H}}_n=0.
\end{equation}
Therefore, we obtain the MMSE receiver for single-stream transmission as
\begin{equation}
  \label{eq_mmse_filter_final}
  \mathbf{u}^*_n=\sqrt{P_n}\left(\sum\limits_{i=1}^{N}P_i\mathbf{H}_{i,n}\mathbf{w}_i\mathbf{w}^{\textrm{H}}_i\mathbf{H}^{\textrm{H}}_{i,n}+\sigma_n^2\mathbf{I}\right)^{-1}\mathbf{H}_{n,n}\mathbf{w}_n.
\end{equation}

Let $\mathbf{R}_n=\sum_{i\ne n}P_i\mathbf{H}_{i,n}\mathbf{w}_i\mathbf{w}^{\textrm{H}}_i\mathbf{H}^{\textrm{H}}_{i,n}+\sigma_n^2\mathbf{I}$ denote the covariance matrix of the interference plus noise at DS $n$. Then, for a given transmit power strategy profile $\mathbf{p}$, the receiving-filter vector at DS $n$ in (\ref{eq_mmse_filter_final}) can be rewritten as
\begin{equation}
  \label{eq_mmse_filter_final_IN}
  \mathbf{u}^*_n=\sqrt{P_n}\left(P_n\mathbf{H}_{n,n}\mathbf{w}_n\mathbf{w}^{\textrm{H}}_n\mathbf{H}^{\textrm{H}}_{n,n}+\mathbf{R}_n\right)^{-1}\mathbf{H}_{n,n}\mathbf{w}_n.
\end{equation}
Observing (\ref{eq_mmse_filter_final_IN}), we note that acquiring the MMSE receiving filter only needs the information about the local channel estimation and the measurement of interference $\mathbf{R}_n$ at the DS side. Therefore, $\mathbf{u}_n$ can be computed independently at each DS as long as the transmit powers and the Tx-beamformers of the links remain unchanged. Furthermore, consider that the QoS of link $n$ is measured as a function of the SINR $\gamma_n$ in (\ref{eq_local_optimal_a}), for instance, as given in (\ref{eq_local_optimal_instance}). Then, by the well-known fact $\textrm{MMSE}_n=E\{\Vert s\Vert^2\}/(1+\gamma_n)$~\cite[(8.121)]{tse2005fundamentals, 4509444}, adopting the MMSE receiving filter in (\ref{eq_mmse_filter_final_IN}) helps to reduce the local transmit power $P_n$ for any given set of Tx-beamformers $(\mathbf{w}_1,\ldots,\mathbf{w}_N)$ at the equality condition of (\ref{eq_local_optimal_a}), see also (\ref{eq_SINR}). Therefore, we have the following theorem:
\begin{Theorem}
  \label{thm_convergence_two_stage}
  Consider a joint power allocation and Rx-beamforming problem, which is reduced from (\ref{eq_local_optimal_instance}) with a fixed set of Tx-beamformers and a QoS constraint as a monotonically increasing function of the local SINR $\gamma_n$ for each $n$. Then, the asynchronous, iterative strategy update of powers and Rx-beamformers by sequentially applying (\ref{eq_mmse_filter_final_IN}) and Algorithm~\ref{alg_jacobian} is guaranteed to converge if a feasible solution exists.
\end{Theorem}
\begin{proof}
For each link $n=\in\mathcal{N}$, we consider two sequences $\{\mathbf{u}_n(t)\}_{t=0}^{\infty}$ and $\{P_n(t)\}_{t=0}^{\infty}$, where $\mathbf{u}_n(t+1)$ is obtained based on (\ref{eq_mmse_filter_final}) with the joint power strategies $(P_n(t), P_{-n}(t))$. Based on the fact that $\textrm{MMSE}_n(t)=1/(1+\gamma_n(t))$, where $\gamma_n(t)$ is independent of $\mathbf{u}_{-n}(t)$ according to (\ref{eq_SINR}), we know that
\begin{equation}
  \label{eq_nonincreasing_SINR}
  \gamma_n(P_n(t), \mathbf{u}_{n}(t+1); P_{-n}(t))\ge \gamma_n(P_n(t), \mathbf{u}_{n}(t); P_{-n}(t)),
\end{equation}
by definition of MMSE. According to (\ref{eq_local_optimal_fixed_bfv}), given the fixed Tx-beamformers $\mathbf{w}_i$ ($i\in\mathcal{N}$), $\mathbf{u}_n(t+1)$ and the adversary power allocation $P_{-n}(t)$, the best response $P_n(t+1)$ in Algorithm~\ref{alg_jacobian} is always obtained when the equality condition of (\ref{eq_local_optimal_fixed_bfv_a}) is met. Namely, the best response of local power strategy is always obtained at the lower bound of the feasible region $\mathcal{P}_n(P_{-n}(t))$, i.e., $\gamma_n(P_n(t), \mathbf{u}_{n}(t); P_{-n}(t)) = \overline{\gamma}_n$. Therefore, we have $\gamma_n(P_n(t), \mathbf{u}_{n}(t+1); P_{-n}(t))\ge \overline{\gamma}_n$, and thus obtain $P_n(t+1)\le P_n(t)$ in the next-round power allocation game. This indicates that the one-step update $\mathbf{u}(t+1)$ always expands the lower bound of the feasible region of the power strategy for link $n$, i.e., $\mathcal{P}_n(P_{-n}(t))$. Then, the iterative strategy update using the MMSE receiver leads to a monotonically deceasing sequence $\{P_n(t)\}_{t=0}^{\infty}$ for all $n\in\mathcal{N}$; hence their convergence, by compactness of $\mathcal{P}_n(P_{-n}(t))$, with Algorithm~\ref{alg_jacobian}. In return, with the mapping between $\mathbf{p}$ and $\mathbf{u}_n$ shown in (\ref{eq_mmse_filter_final}), the convergence of $\{\mathbf{u}_n(t)\}_{t=0}^{\infty}$ ($n\in\mathcal{N}$) is also guaranteed, since $\forall n\in\mathcal{N}$, $\{\textrm{MMSE}_n(t)\}_t$ is non-increasing and lower-bounded by zero.
\end{proof}

For SINR-based QoS constraints, based on Theorem~\ref{thm_convergence_two_stage}, we are able to develop the two-stage joint power allocation and Tx-Rx beamformer control scheme in Algorithm~\ref{alg_two_stage}. Therein, the power allocation and Rx-beamforming strategy is obtained based on the two-stage iteration which alternates between the non-cooperative power allocation game and MMSE Rx-beamforming (see Lines 3 and 5). We consider that no information about the interference channel $\mathbf{H}_{i,n}$ ($i\ne n$) is available to link $n$. Then, the most straightforward approach for attaining a locally optimal Tx-beamforming vector $\mathbf{w}^*_n$ is to treat the interference from the other links as stationary noises, and maximize the perceived signal-to-noise ratio (see also (\ref{eq_SINR})) by solving $\max_{\mathbf{w}_n:\Vert\mathbf{w}_n\Vert^2=1}\Vert\mathbf{u}_n^{\textrm{H}}\mathbf{H}_n\mathbf{w}_n \Vert^2$, hence (\ref{eq_matched_filter}). This leads to the well-known form of the matched filter for a single link~\cite{6832894}, and the myopic solution $\mathbf{w}^*_n$ is obtained as the eigenvector corresponding to the maximum eigenvalue of $\mathbf{H}^{\textrm{H}}_n\mathbf{u}_n\mathbf{u}^{\textrm{H}}_n\mathbf{H}_n$.
\setcounter{algorithm}{1}
\begin{algorithm}[t]
    \begin{small}
 \caption{Two-stage search for joint power allocation and Tx-Rx beamforming}
 \begin{algorithmic}[1]
 \REQUIRE
 $\forall n\in\mathcal{N}$: select ${P}_n=\overline{P}_n$. $\forall n\in\mathcal{N}$: initialize $\mathbf{u}_n$ and $\mathbf{w}_n$ randomly. Select $\varepsilon>0$.
 \REPEAT
   \STATE  $\forall n\in\mathcal{N}$: $P'_n\leftarrow P_n$,  $\mathbf{u}'_n\leftarrow\mathbf{u}_n$ and $\mathbf{w}'_n\leftarrow\mathbf{w}_n$
   \STATEx Stage 1:
   \STATE  $\forall n\in\mathcal{N}$: update $P_n$ following Algorithm~\ref{alg_jacobian} with $\mathbf{u}_i$ and $\mathbf{w}_i$ ($i\in\mathcal{N}$)
   \STATEx Stage 2:
   \FORALL{$\forall n\in\mathcal{N}$}
        \STATE Update $\mathbf{u}_n$ according to (\ref{eq_mmse_filter_final_IN}) with ${P}_i$, $\mathbf{u}'_i$ and $\mathbf{w}'_i$ ($i\in\mathcal{N}$)
        \STATE Update $\mathbf{w}_n$ only once with the myopic matched filter scheme as
        \begin{equation}
          \label{eq_matched_filter}
          \mathbf{w}_n=\arg\max_{\mathbf{w}_n:\Vert\mathbf{w}_n\Vert^2=1}\Vert\mathbf{u}_n^{\textrm{H}}\mathbf{H}_n\mathbf{w}_n \Vert^2.
        \end{equation}
   \ENDFOR
 \UNTIL {$\forall n\in\mathcal{N}$: $\Vert \mathbf{u}_n-\mathbf{u}'_n\Vert\le\varepsilon$ {\bf{and}} $\Vert \mathbf{w}_n-\mathbf{w}'_n\Vert\le\varepsilon$ {\bf{or}} $\exists P_n\in\mathcal{N}$, $P_n$ is infeasible in Stage 1}
 \end{algorithmic}
 \label{alg_two_stage}
\end{small}
\end{algorithm}

It is worth noting that other iterative Tx-beamforming schemes can be adopted to replace the myopic matched filter in Line 6 of Algorithm~\ref{alg_two_stage}. According to (\ref{eq_SINR}), an update in the Tx-beamformer $\mathbf{w}_n$ of SS-DS link $n$ will also influence its interference to the other links. Theoretically, it is necessary to examine the impact of the levels of available link information (e.g., estimates of mutual interference) at each link on the performance of the Tx-beamforming scheme. Fortunately, we observe in simulations (see Section~\ref{Sec:Simulation}) that different iterative Tx-beamforming schemes, which are based on different levels of shared CSI information about the interference links, provide limited improvement to the performance of the network. Nevertheless, theoretical discussion about the impact of information exchange on the algorithm convergence is still necessary. Therefore, we provide in Appendix~\ref{lab_appendix} the design of a series of possible ad-hoc iterative methods to optimize the Tx-beamformers according to different levels of information sharing regarding the interference link state information among the local links.

% Note the simulation results are replaced by the new ones, due to the changes in the function  of supply power (see Eq. 6)
\section{Simulation Results}\label{Sec:Simulation}
\begin{figure*}[t!]
\centering     %%% not \center
\subfigure[]{\label{fig_performance_mesh_a}\includegraphics[width=.30\linewidth]{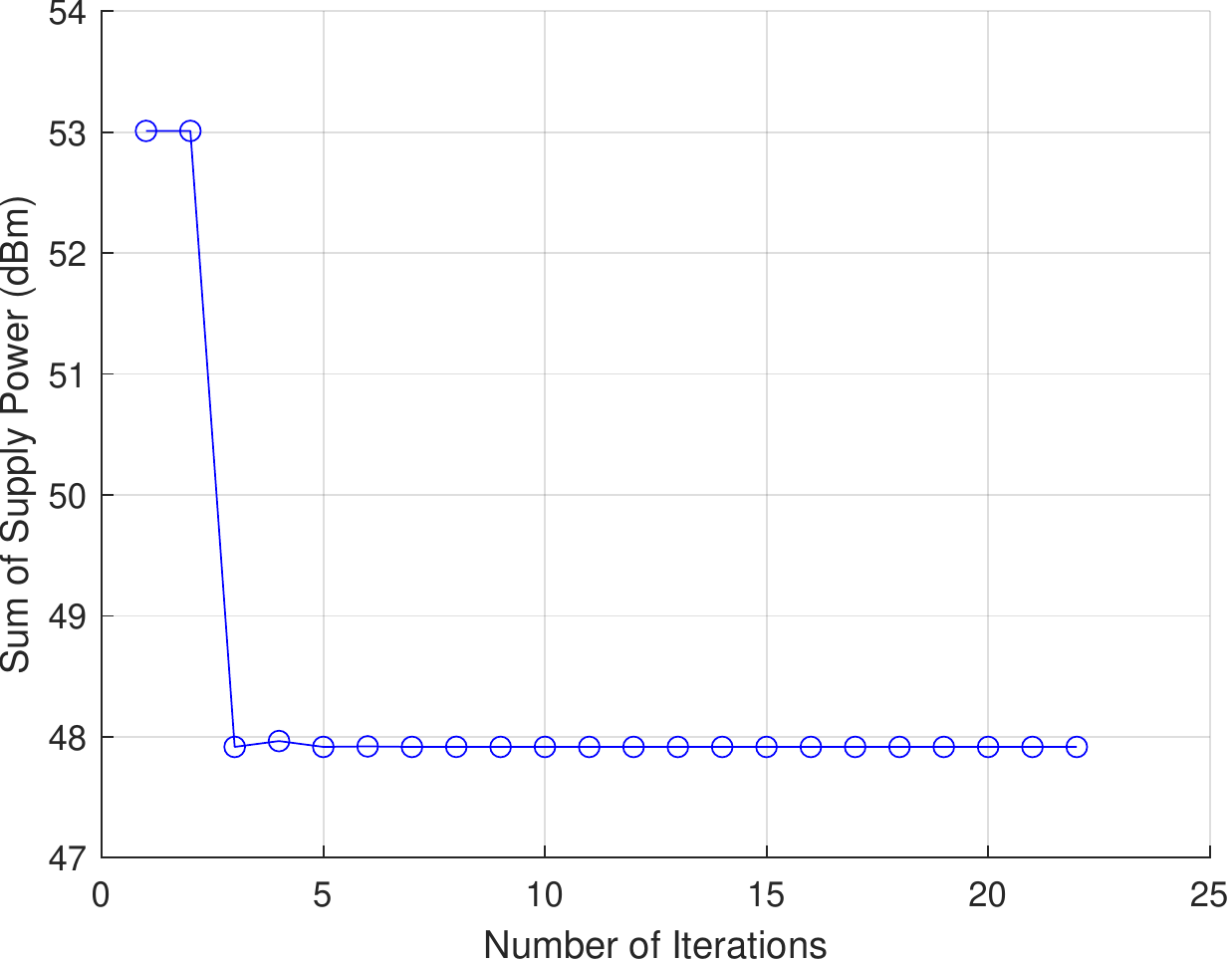}}
\subfigure[]{\label{fig_performance_mesh_b}\includegraphics[width=.34\linewidth]{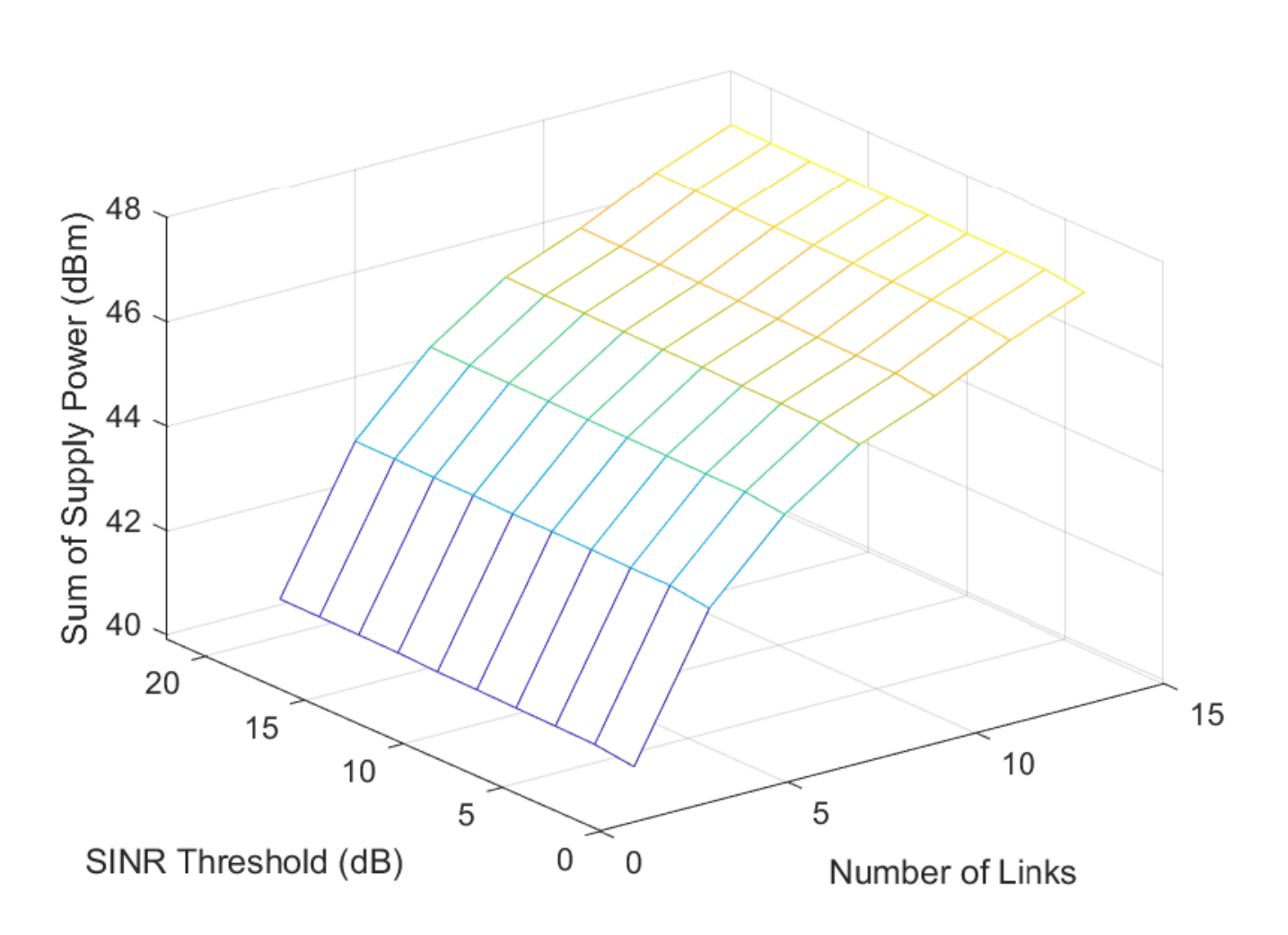}}
\subfigure[]{\label{fig_performance_mesh_c}\includegraphics[width=.34\linewidth]{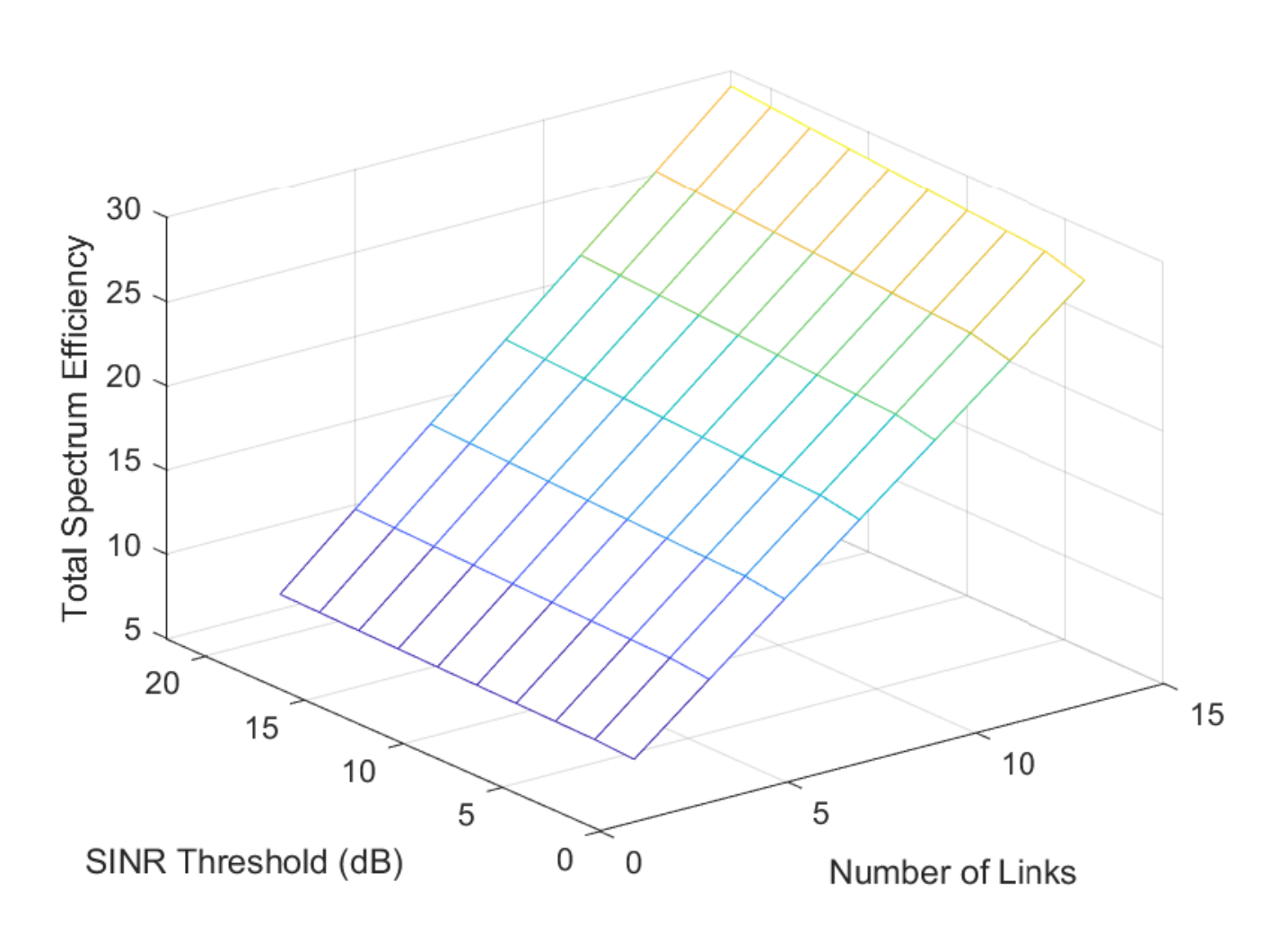}}
\caption{Monte Carlo simulation results for Algorithm~\ref{alg_two_stage} using matched filters for Tx-beamforming.  (a) A snapshot of the power evolution with Algorithm~\ref{alg_two_stage} for 10 links and SINR thresholds $\overline{\gamma}_n=20$dB.
(b) Sum of supply power vs. different numbers of links and SINR thresholds. (c) Sum of spectrum efficiency vs. different link numbers and SINR thresholds.}
\label{fig_performance_mesh}
\end{figure*}

Considering the characteristics of the mmWave band, we use an urban outdoor environment with non-LoS components channel model for our simulations. Since scattering in mmWave bands is limited, multipath is mainly caused by reflections~\cite{7390101}. Hence, mmWave channels are determined by the geometry of the relevant antenna arrays. Hence in this section, we adopt the clustered channel model for the link between each SS $i$ and any DS $n$ from~\cite{6717211} as follows:
\begin{align}
    \label{eq_channel_model}
     \mathbf{H}_{i,n}= &\beta\sum\limits_{j,l}\alpha^{j,l}_{i,n}\mathbf{g}_{i,n}(\phi_{\textrm{r},n}^{j,l},\theta_{\textrm{r},n}^{j,l}, \phi_{\textrm{t},i}^{j,l},\theta_{\textrm{t},i}^{j,l})\nonumber \\
    & \mathbf{a}_{\textrm{r},n}(\phi_{\textrm{r},n}^{j,l},\theta_{\textrm{r},n}^{j,l}) \left(\mathbf{a}_{\textrm{t},i}(\phi_{\textrm{t},i}^{j,l},\theta_{\textrm{t},i}^{j,l})\right)^*,
    %\mathbf{g}^r_n(\phi^r_{j,l},\theta^r_{j,l})\mathbf{g}^t_i(\phi^t_{j,l},\theta^t_{j,l})
\end{align}
where $\beta=\sqrt{KL/N^{\textrm{cl}}_{i,n}N^{\textrm{ray}}_{i,n}}$ is a normalized factor, $K$ and $L$ are the number of antennas at SS $i$ and DS $n$, $N^{\textrm{cl}}_{i,n}$ is the number of scattering visible clusters and $N^{\textrm{ray}}_{i,n}$ is the number of propagation paths in each cluster. $\alpha^{j,l}_{i,n}$ is the complex pathloss coefficient of the $l$-th ray in the $j$-th scattering cluster and depends on the Tx-Rx distance between antennas along the corresponding path. In (\ref{eq_channel_model}), $\phi_{\textrm{r},n}^{j,l}$ (cf. $\theta_{\textrm{r},n}^{j,l}$) and $\phi_{\textrm{t},i}^{j,l}$ (cf. $\theta_{\textrm{t},i}^{j,l}$) are the corresponding random central azimuth (elevation) angles of arrival (i.e., at DS $n$) and departure (i.e., at SS $i$), respectively.
$\mathbf{g}_{i,n}(\phi_{\textrm{r},n}^{j,l},\theta_{\textrm{r},n}^{j,l}, \phi_{\textrm{t},i}^{j,l},\theta_{\textrm{t},i}^{j,l})$ is the $(j,l)$-th element complex gain between the transmit and the receive antennas from SS $i$ to DS $n$ at the corresponding angles of arrival and departure.
%$\mathbf{g}^r_n(\phi^r_{j,l},\theta^r_{j,l})$ and $\mathbf{g}^t_i(\phi^t_{j,l},\theta^t_{j,l})$ are the element gains of the receive and transmit antennas at the corresponding angles of arrival and departure.
$\mathbf{a}_{\textrm{r},n}(\phi_{\textrm{r},n}^{j,l},\theta_{\textrm{r},n}^{j,l})$ and $\mathbf{a}_{\textrm{t},i}(\phi_{\textrm{t},i}^{j,l},\theta_{\textrm{t},i}^{j,l})$ are the normalized receive and transmit array response vectors at the azimuth (elevation) angle of $\phi_{\textrm{r},n}^{j,l}$ ($\theta_{\textrm{r},n}^{j,l}$) and $\phi_{\textrm{t},i}^{j,l}$ ($\theta_{\textrm{t},i}^{j,l}$), respectively.

For the purpose of illustration, we randomly place each SS/DS in the plane with roughly equal distance between the neighboring stations. We ensure that the inter-link interference is mainly caused by  tier-1 neighbor stations of each DS, such that a sufficiently good SINR (e.g., 20dB) can be achieved at each link. We set the number of antennas of both the SS and the DS to be $8$, and the spectrum efficiency of link $n$ is measured as $\log(1+\gamma_n)$. In our first experiment, in Figure~\ref{fig_performance_mesh}, we adopt the myopic matched filter for Tx-beamforming, and demonstrate the efficiency of the proposed joint power allocation and beamforming algorithm (i.e., Algorithm~\ref{alg_two_stage}) in terms of convergence speed (Figure~\ref{fig_performance_mesh_a}) and overall network performance (Figures~\ref{fig_performance_mesh_b} and~\ref{fig_performance_mesh_c}). Since the varying QoS requirements along the ``SINR Threshold (dB)'' axis is met by all the links, the total spectrum efficiency of the network increases linearly w.r.t. the number of operating links in Figure~\ref{fig_performance_mesh_c}. On the other hand, Figure~\ref{fig_performance_mesh_b} indicates that the size of the network has a larger impact on the sum of supply powers than the levels of the SINR threshold.

Furthermore, we introduce three baseline algorithms for joint power allocation and beamforming performance comparison with the proposed Algorithm~\ref{alg_two_stage}. The first baseline algorithm, ``Global Iterative Zero Forcing (ZF)'', is adapted from~\cite{6760591} to iteratively enforce both the Tx and the Rx filters of each link to be in the null space of the equivalent interfering channels with a modified SINR constraint\footnote{Exchanges CSI among links are assumed for global iterative ZF. Instead of integrating ZF into the MMSE-based Rx-beamforming procedure~\cite{6760591}, the adopted scheme uses stand-alone ZF for both Tx- and Rx-beamforming.}. The second baseline algorithm, ``Global Coordinated Tx-BeamForming (BF)'', is adopted from~\cite{5463229} as a centralized Tx-beamforming procedure for sum-of-power minimization with SINR constraints through full link coordination using fixed Rx beamforming vectors. Likewise, it is assumed that the CSI of all interfering links is known to each transmitter.
The third algorithm is a global iterative Rx-Tx coordinated MSE beamformer, ``Global Coordinated-MSE Tx-Rx BF'' where the transmit beamformers use MSE beamformers based on full CSI following (\ref{eq_local_optimum_TxBF_multiplier_modified}).

\begin{figure*}[t!]
\centering     %%% not \center
\subfigure[]{\label{fig_performance_vs_linknumber_a}\includegraphics[width=.36\linewidth]{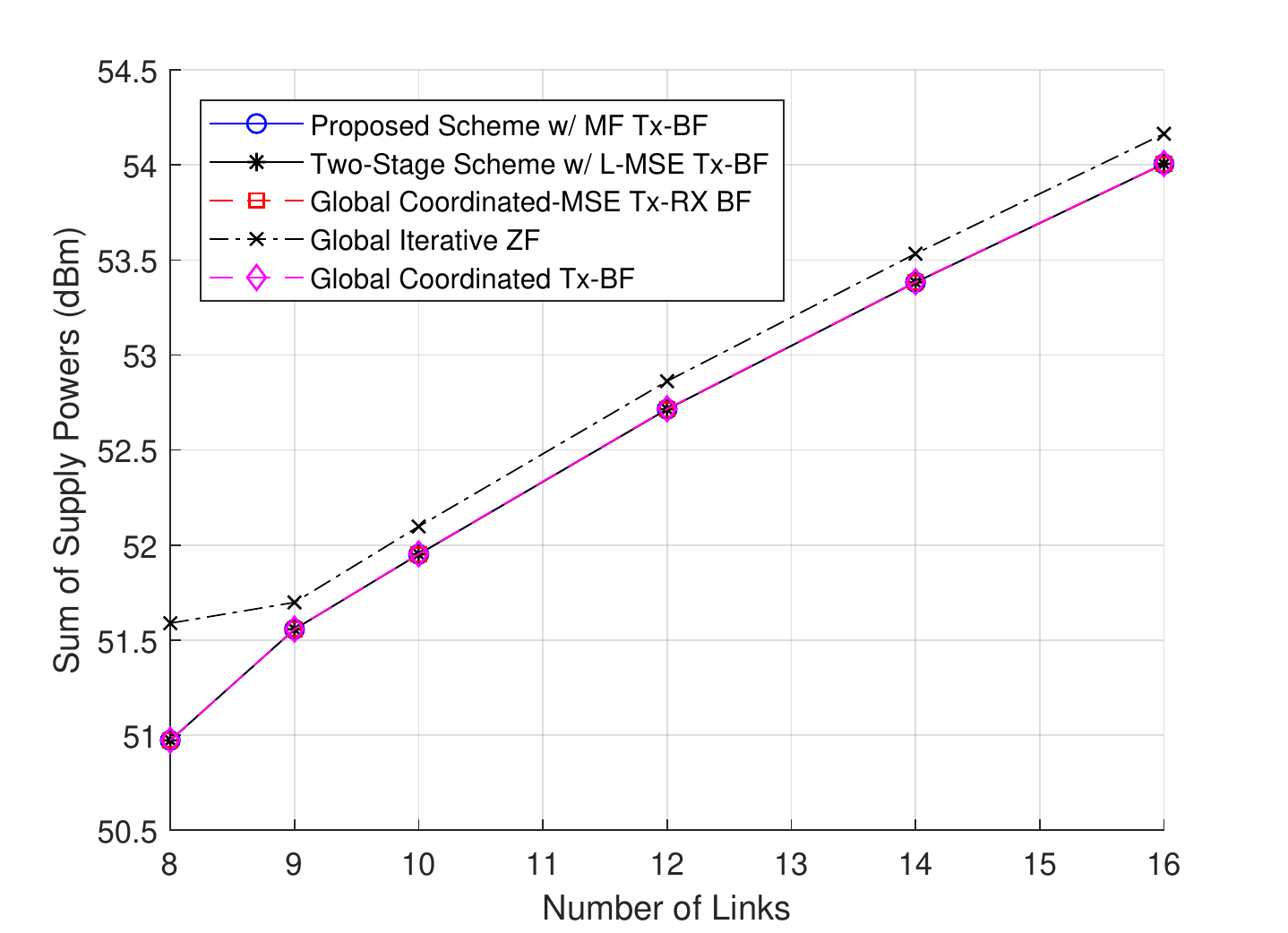}}
\subfigure[]{\label{fig_performance_vs_linknumber_b}\includegraphics[width=.36\linewidth]{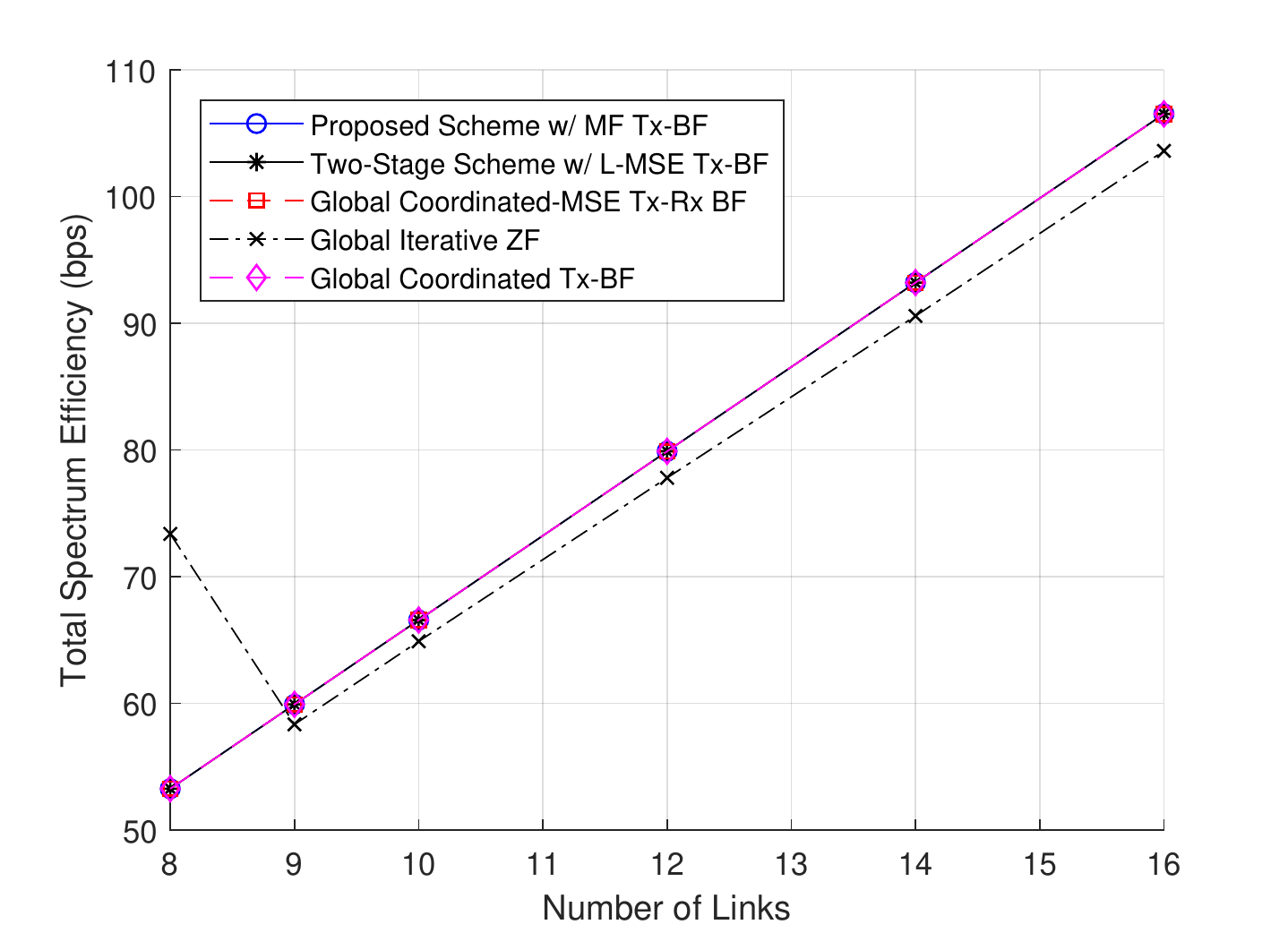}}
\caption{Performance comparison for different numbers of links through Monte Carlo simulations at an SINR threshold of 20dB for all the links. (a) Sum of supply power vs. numbers of links for different algorithms. (b) Sum of spectrum efficiency vs. numbers of links for different algorithms.}
\label{fig_performance_vs_linknumber}
\end{figure*}

To show that using a Tx-beamformer matched to the channel is sufficient, we also show by simulation that using an MMSE based Tx-beamformer does not reduce the power consumption. This scheme appears in the simulations as: ``Two-Stage Scheme with Local-MSE (L-MSE) Tx-BF'' (see (\ref{eq_local_optimum_TxBF_multiplier})). As shown by the four overlapping curves in Figure~\ref{fig_performance_vs_linknumber_a} and Figure~\ref{fig_performance_vs_linknumber_b}, the proposed joint allocation framework achieves the same level of performance as the optimal fully-coordinated algorithm, ``Global Coordinated-MSE Tx-Rx BF'' and as optimizing only the Tx beamformers with a fixed Rx beamformer (i.e., ``Global Coordinated Tx-BF''). As discussed in the paper, the reason for the success of the proposed technique is that both the Rx beamformer adaptation and the power adaptation stages gradually improve the interference environment for all stations.

\begin{figure*}[t!]
\centering     %%% not \center
\subfigure[]{\label{fig_performance_vs_linknumber_c}\includegraphics[width=.36\linewidth]{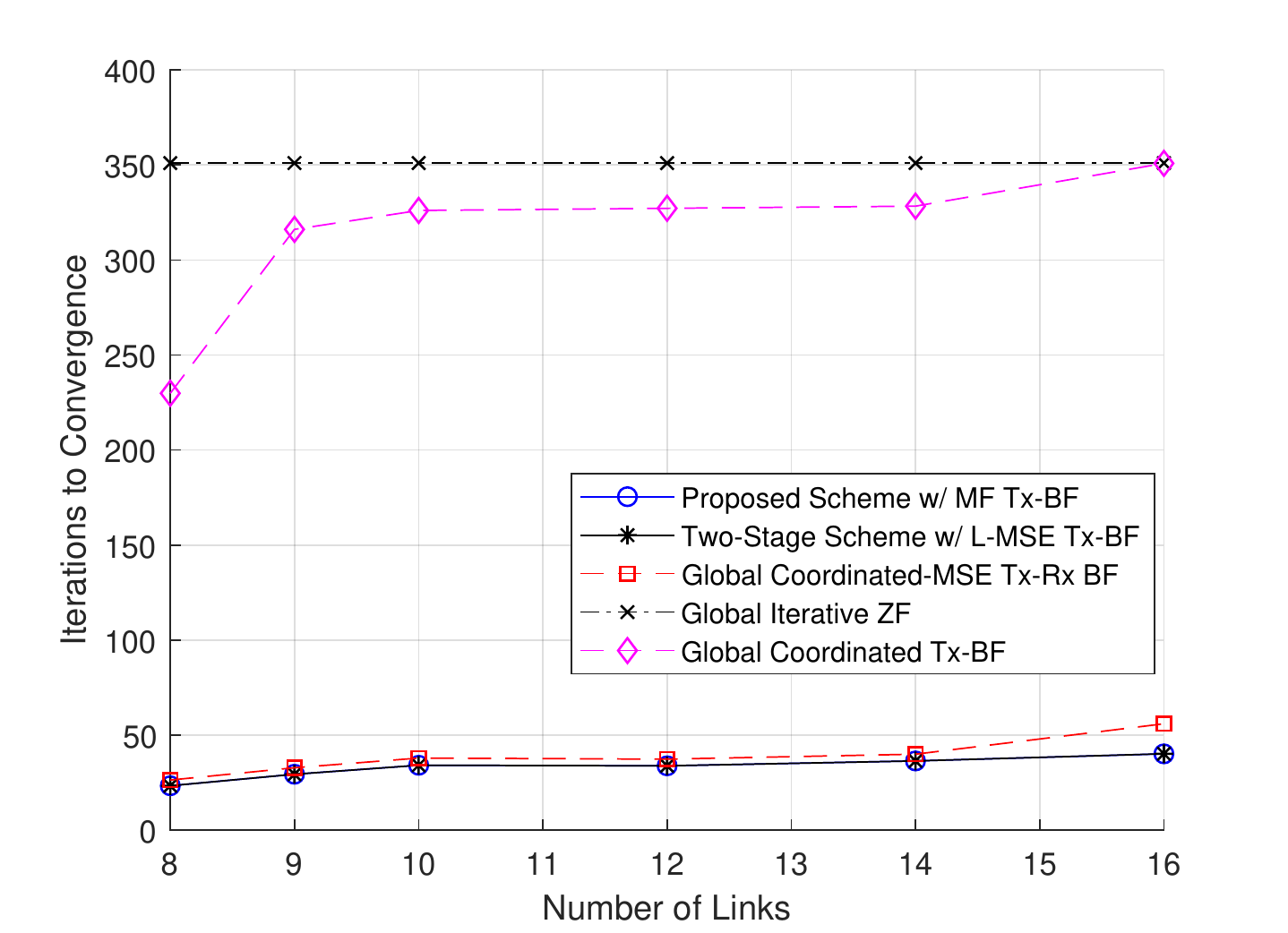}}
\subfigure[]{\label{fig_performance_vs_linknumber_d}\includegraphics[width=.36\linewidth]{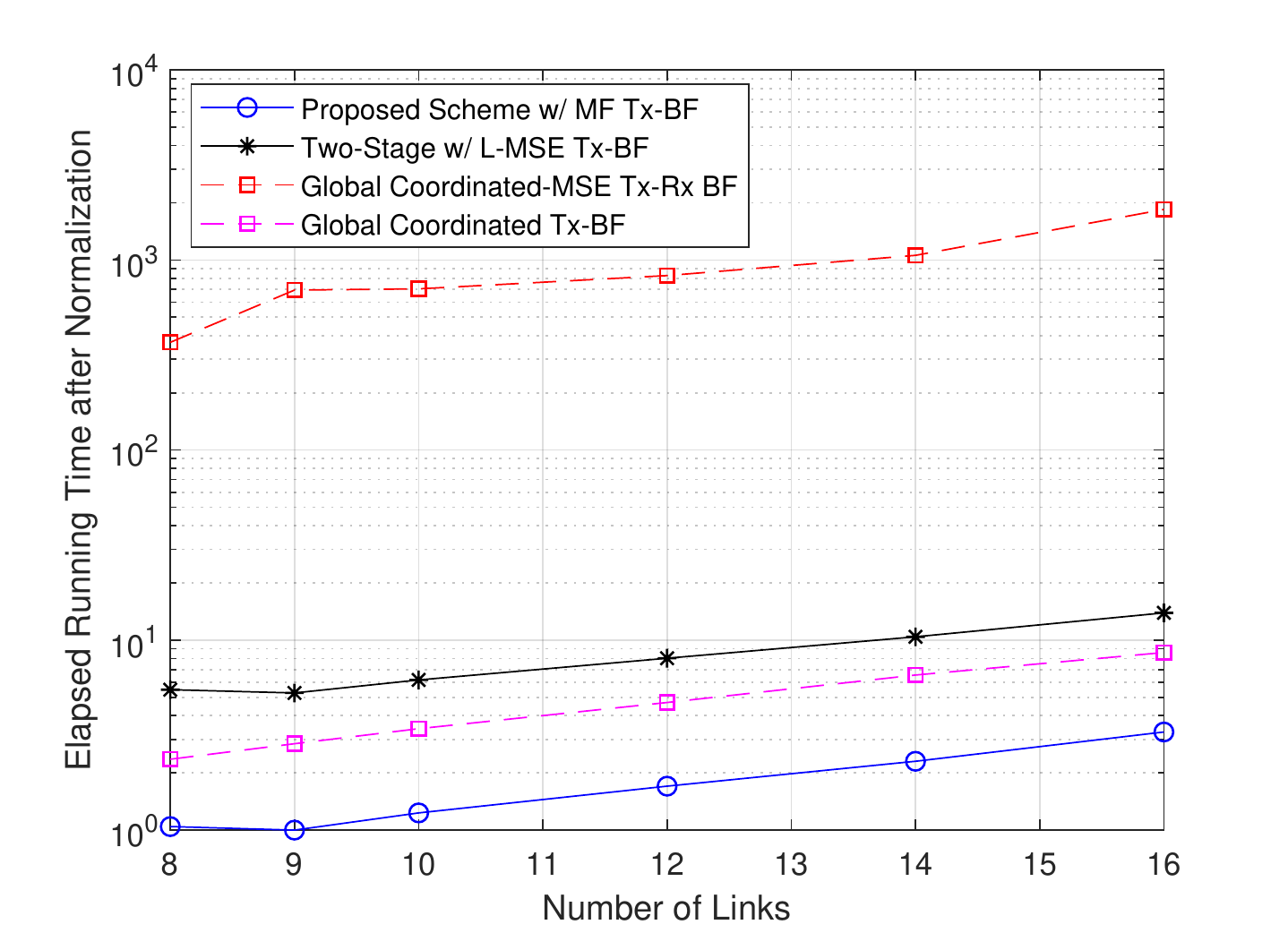}}
\caption{Complexity comparison for different numbers of links through Monte Carlo simulations at an SINR threshold of 20dB for all the links. (a) Number of iterations to convergence vs. numbers of links for different algorithms. (b)  Elapsed running time to convergence (after normalization w.r.t. the time used by the MF Tx-BF algorithm for a network of 5 links) vs. numbers of links for different algorithms.}
\label{fig_time_complexity_vs_linknumber}
\end{figure*}

Figure~\ref{fig_time_complexity_vs_linknumber} depicts both the number of iterations required by each algorithm as well as the overall simulation run time. We can clearly see that as the network size increases, the ``Global Coordinated Tx-BF'' and the ``Global Coordinated-MSE Tx-Rx BF'' schemes both require significantly more iterations to reach the same level of performance (i.e., convergence) compared to the proposed algorithm (see Figure~\ref{fig_performance_vs_linknumber_c}).  Since the proposed scheme requires $O(N)$ signals per iteration, while the full CSI schemes require $O(N^2)$ messages per iteration, we observe that the proposed scheme has approximately, factor of $7N$ less messages required. Moreover, it does not require a complicated estimation of the channel between all transmitters and receivers. It is interesting to observe that experimentally the number of total power updates grow only linearly with the number of links.

\begin{table} [!t]
    \caption{Average number of power iterations till convergence by the proposed scheme vs. number of links.}
  \centering
 \begin{tabular}{ |p{2cm}|c|c|c|c|c|c|}
  \hline
  No. of Links &  8 & 9 & 10 & 12 & 14 & 16 \\
  \hline
  Average No. of Power Iterations & 40.4
   & 45.1 &   47.2 &   56.7 &   58.3 & 65.6\\
  \hline
\end{tabular}
  \label{Number_Power_Iterations_For_Response}
\end{table}

In addition to Figure~\ref{fig_performance_vs_linknumber_c}, Table~\ref{Number_Power_Iterations_For_Response} provides an empirical study of the number of power adaptation messages during the running of the algorithm with Monte-Carlo simulations. Using linear regression we obtain that the number of messages fits very well to $16+3N$, where $N$ is the number of links. For notification of non-terminating iteration by each link in Algorithm 1, the signaling complexity is trivially bounded $O(N)$. In contrast, any method which requires knowledge of {\em all} the channel coefficients between any receiver and any transmitter is $O(N^2)$. Also note, that channel estimation, of all these channels requires more complicated training scheme for channel estimation as well as close coordination between transmitters.

In Figure~\ref{fig_performance_vs_linknumber_d}, we further examine the computational complexity of the various schemes by comparing the average CPU time consumed by each scheme. The elapsed running time to convergence for the three reference schemes is normalized w.r.t. that of the proposed ``MF Tx-BF''. As we can observe in Figure~\ref{fig_performance_vs_linknumber_d}, ``Global Coordinated-MSE Tx-Rx BF'' has the highest time complexity, since no closed-form solution is accessible and a numerical convex optimization solver (i.e., CVX) is needed for solving (\ref{eq_local_optimum_TxBF_multiplier_modified}). Comparatively, at each iteration, ``MF Tx-BF'', ``L-MSE Tx-BF'' and ``Global Coordinated Tx-BF'' are derived using closed-form solutions and thus have a similar level of time complexity in Algorithm~\ref{alg_two_stage}. Interestingly, the proposed scheme has the lowest time complexity (at least 3 times faster than any of the other schemes).

\begin{figure}[t]
  \centering
  \includegraphics[width=0.38\textwidth]{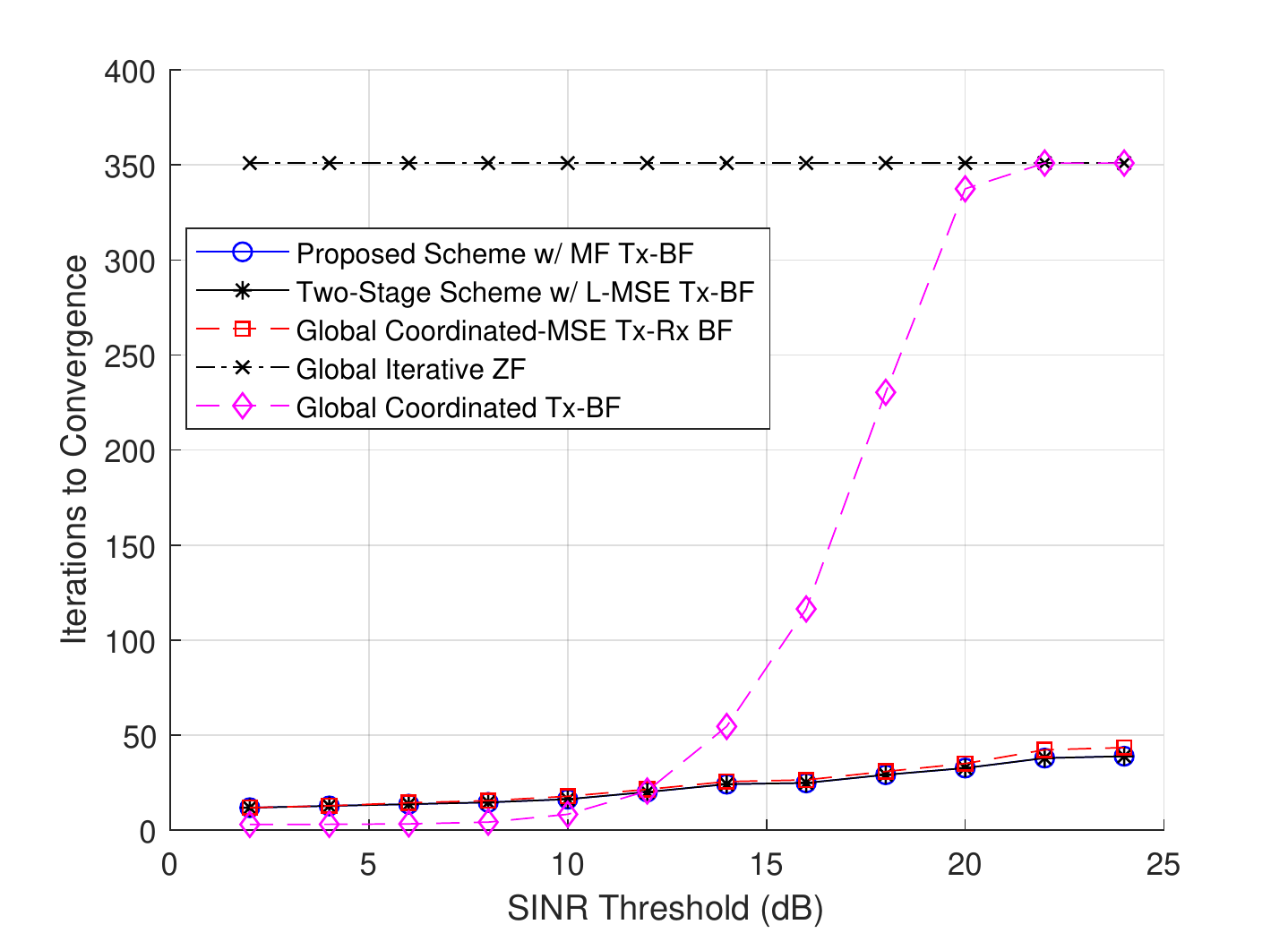}
  \caption{Monte-Carlo simulations for number of iterations to convergence vs. SINR thresholds.}
  \label{figure_IterationvsSINR}
\end{figure}

\begin{figure}[!t]
  \centering
  \includegraphics[width=0.38\textwidth]{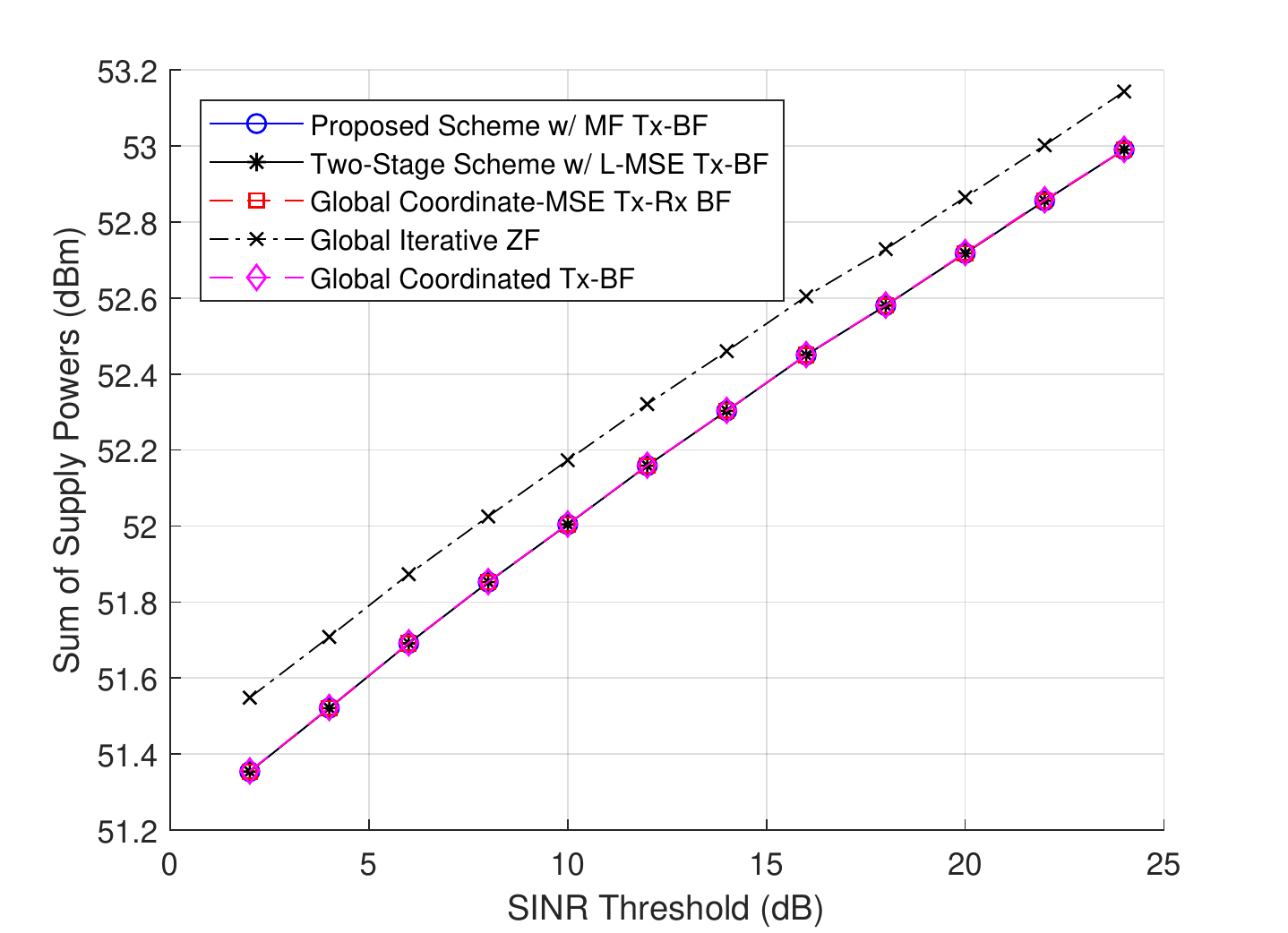}
  \caption{Monte-Carlo simulations for total supply power vs. SINR thresholds. The proposed scheme with three different Tx-BF methods and the Global Coordinated Tx-BF scheme achieve the same level of energy efficiency.}
  \label{figure_SupplyPowervsSINR}
\end{figure}

In the last experiment (see Figures~\ref{figure_IterationvsSINR} and~\ref{figure_SupplyPowervsSINR}), we set the number of operating links to be 14 and compare the performance of the algorithms as a function of the SINR thresholds (assuming the same required SINR level on all links). Figure~\ref{figure_IterationvsSINR} demonstrates that for practical target SINR levels above 10 dB the proposed method is significantly better. Figure~\ref{figure_SupplyPowervsSINR} presents the total required power for achieving the target SINR. We can clearly see that the proposed method yields identical performance to the coordinated techniques.

\section{Conclusion}\label{Sec:Conclusion}
In this paper we analyzed the existence and convergence of generalized Nash equilibrium in monotonic generalized games. Then we exploited the properties of monotonic games to derive a two stage algorithm for joint beamforming and power allocation under SINR constraints. We showed that the scheme always converges. Furthermore, we showed through simulations that with a matched filter beamformer at the transmitter we are able to achieve optimal performance comparable to the global optimum. Moreover, the proposed technique requires significantly less signaling (by a factor of $7N$ and has significantly lower computational complexity.

The results are applicable to ad-hoc MIMO networks under general QoS constraints (not necessarily convex), which correspond to the practical scenario of wireless backhaul communications in mmWave networks. Recently, we also showed that the proposed framework of monotonic games for power allocation can be used to optimize channel aware, energy efficient Aloha networks \cite{wenbo_icc_2022}.
\appendices
\section{Alternative Example  of Generalized Nash Games with Non-Convex QoS Constraints}
\label{app_non_convex_example}
In comparison to the the generalized problem in (\ref{eq_local_optimal_instance}) based on the assumption of perfect CSI, we can also consider an alternative case where only Channel Distribution Information (CDI) is available to the SS-DS links. Namely, the channel matrices are assumed to be random variables drawn from a complex Gaussian distribution, $\textrm{vec}(\mathbf{H}_{i,n})\sim\mathcal{CN}(0,\mathbf{\Sigma}_{i,n})$. Then in (\ref{eq_local_optimal_instance}), the local optimization problem based on the constraint of instantaneous SINR can be replaced by
\begin{subequations}\label{eq_local_optimal_instance_2}
\begin{align}
  \tag{\ref{eq_local_optimal_instance_2}}
    & (P^*_n, \mathbf{u}^*_n, \mathbf{w}^*_n) = \arg \min\limits_{P_n, \mathbf{u}_n, \mathbf{w}_n}  c_n(P_n) \\
        \label{eq_local_optimal_a_instance_2}
    \textrm { s.t. } & \Pr(\gamma_n(P_n, \mathbf{u}_n, \mathbf{w}_n, P_{-n}, \mathbf{u}_{-n}, \mathbf{w}_{-n})\ge\underline{\gamma}_n)\ge \overline{q}_n, \\
    \label{eq_local_optimal_b_instance_2}
    & \underline{P}_n\le P_n\le \overline{P}_n, \\
    \label{eq_local_optimal_c_instance_2}
    & \Vert \mathbf{w}_n \Vert = 1,
\end{align}
\end{subequations}
where (\ref{eq_local_optimal_a_instance_2}) ensures that the probability of the local SINR being above a certain threshold $\underline{\gamma}_n$ is greater than a given QoS threshold $\underline{q}_n$, with $(1-\overline{q}_n)$ indicating the link outage probability.

Assume that all SS-DS links experience correlated Rayleigh fading. Then by~\cite{5580109}[Theorem 1], the QoS constraint function on the left-hand side of (\ref{eq_local_optimal_a_instance_2}) can be expressed as
\begin{align}
  \label{eq_alternative_qos}
  q_n(\gamma_n(P_n, \mathbf{u}_n, \mathbf{w}_n, P_{-n}, \mathbf{u}_{-n}, \mathbf{w}_{-n})) = \nonumber\\
  \exp(-\frac{\underline{\gamma}_n}{2}\frac{\sigma^2_n}{g_{n,n}P_n})\prod_{i\ne n}\left(1+\underline{\gamma}_n\frac{g_{i,n}P_i}{g_{n,n}P_n}\right)^{-1},
\end{align}
where $g_{i,n}=(\mathbf{w}_i^*\otimes\mathbf{u}_n)^{\textrm{H}}\mathbf{\Sigma}_{i,n}(\mathbf{w}_i^*\otimes\mathbf{u}_n)$, with $\otimes$ representing the operation of Hadamard product. With fixed sets of $\mathbf{w}_n$ and $\mathbf{u}_n$ for all $n\in\mathcal{N}$, we have $q(P_n,P_{-n})$ in (\ref{eq_alternative_qos}) being monotonically increasing in $P_n$ and decreasing in $P_i$ for all $i\ne n$. Then, the non-convex power allocation problem obtained from (\ref{eq_local_optimal_instance_2}) also fits into our proposed framework of monotonic generalized Nash games, and can be solved in a distributed manner following Algorithm~\ref{alg_jacobian}\footnote{For other examples of employing monotonic generalized Nash games for decentralized optimization in wireless networks, we refer the readers to our recent work~\cite{wenbo_icc_2022}.}.

\section{Tx-Beamforming Based on Different Levels of Shared CSI Information}\label{lab_appendix}
\subsection{Tx-Beamforming Based on Local CSI}
Inspired by the derivation of the myopic matched filter and following our discussion regarding (\ref{eq_mse}), it is natural to obtain another Tx-beamformer (cf. Line 6 in Algorithm~\ref{alg_two_stage}) based solely on the local CSI as $\mathbf{w}^*_n =\arg\min_{\mathbf{w}_n:\Vert\mathbf{w}_n\Vert^2=1}\left\{\textrm{MSE}_n\right\}$. By relaxing the constraint of the normalized Tx-beamformer, we are able to obtain the following sub-problem of quadratic programming to determine $\mathbf{w}_n$:
\begin{equation}
  \label{eq_optimizing_TxBF}
  \mathbf{w}^*_n =\arg\min\limits_{\mathbf{w}_n}E\left\{\Vert\hat{s}_n-s_n\Vert^2\right\} \textrm{ s.t. } \mathbf{w}^{\textrm{H}}_n\mathbf{w}_n\le P_n,
\end{equation}
where $P_n$ is given by the solution of the power allocation game in Algorithm~\ref{alg_jacobian} (cf. Line 3 in Algorithm~\ref{alg_two_stage}). It is straightforward to verify that the strong duality holds for the Lagrange dual problem for (\ref{eq_optimizing_TxBF}). Then, by introducing the Lagrange multiplier $\lambda_n\ge0$ and with (\ref{eq_mse_expand}), we need $\partial (\textrm{MSE}_n+\lambda_n(\mathbf{w}^{\textrm{H}}_n\mathbf{w}_n-P_n))/\partial \mathbf{w}_n=0$ for solving (\ref{eq_optimizing_TxBF}). With a similar technique to the derivation of (\ref{eq_mmse_filter}), we obtain
\begin{equation}
  \label{eq_local_optimum_TxBF_multiplier}
  \mathbf{w}^*_n =\sqrt{P_n}\left(P_n \mathbf{H}^{\textrm{H}}_{n,n}\mathbf{u}_n\mathbf{u}^{\textrm{H}}_n\mathbf{H}_{n,n} + \lambda_n\mathbf{I}\right)^{-1}\mathbf{H}^{\textrm{H}}_{n,n}\mathbf{u}_n,
\end{equation}
where $\lambda_n$ is chosen to satisfy the power constraint in (\ref{eq_optimizing_TxBF}). $\mathbf{w}^*_n$ and $\lambda_n$ can be efficiently obtained with off-the-shelf numerical solvers such as CVX~\cite{cvx2014}. Furthermore, for the first-order optimality condition in (\ref{eq_optimizing_TxBF}) to be met, we exploit the Hermitian property of $P_n\mathbf{H}^{\textrm{H}}_{n,n}\mathbf{u}_n\mathbf{u}^{\textrm{H}}_n\mathbf{H}_{n,n}=\mathbf{V}_n\mathbf{M}_n \mathbf{V}_n^{\textrm{H}}$. By substituting (\ref{eq_local_optimum_TxBF_multiplier}) into the equality constraint condition $\mathbf{w}^{\textrm{H}}_n\mathbf{w}_n=P_n$, we aim to extract the original beam pattern s.t. $\mathbf{w}^{\textrm{H}}_n\mathbf{w}_n=1$ after normalization and obtain
  \begin{align}
    1 =& \tr\left(\mathbf{u}^{\textrm{H}}_n\mathbf{H}_{n,n}\mathbf{V}_n(\mathbf{M}_n+\lambda_n\mathbf{I})^{-2}\mathbf{V}_n^{\textrm{H}}\mathbf{H}_{n,n}^{\textrm{H}}\mathbf{u}_n\right) \nonumber \\
    \label{eq_mse_expand_TxBF}
    = &\tr\left((\mathbf{M}_n+\lambda_n\mathbf{I})^{-2} \mathbf{V}_n^{\textrm{H}}\mathbf{H}_{n,n}^{\textrm{H}}\mathbf{u}_n\mathbf{u}^{\textrm{H}}_n\mathbf{H}_{n,n}\mathbf{V}_n\right) \nonumber \\
    = &\sum\limits_{i=1}^{K}{g_i}/{(\mu_i + \lambda_n)^2},
  \end{align}
where $\mu_i + \lambda_n$ is the $i$-th diagonal element of the diagonal matrix $\mathbf{M}_n+\lambda_n\mathbf{I}$ and $g_i$ is the $i$-th diagonal element of the matrix $\mathbf{V}_n^{\textrm{H}}\mathbf{H}_{n,n}^{\textrm{H}}\mathbf{u}_n\mathbf{u}^{\textrm{H}}_n\mathbf{H}_{n,n}\mathbf{V}_n$. Then, for the execution of Line~6 in Algorithm~\ref{alg_two_stage}, $\mathbf{w}_n$ can be updated with normalization following (\ref{eq_local_optimum_TxBF_multiplier}) after solving (\ref{eq_mse_expand_TxBF}) for $\lambda_n$.

Theoretically, unilaterally adapting $\mathbf{w}_n$ at SS $n$ with respect to the local SINR following (\ref{eq_local_optimum_TxBF_multiplier}) may increase the interference perceived by another link $i$, i.e., ${R}_{n,i}\!=\!P_n\Vert\mathbf{u}^{\textrm{H}}_i\mathbf{H}_{n,i}\mathbf{w}_n\Vert^2$. In this case, the monotonicity of the power-updating sequence $\{P_i(t)\}_{t=0}^{\infty}$ may no longer be guaranteed for link $i$ with Algorithm~\ref{alg_two_stage} if the Tx-beamforming scheme in (\ref{eq_local_optimum_TxBF_multiplier}) is adopted. Although our numerical simulations in Section~\ref{Sec:Simulation} show that with sufficiently good channel conditions, the impact of Tx-beamformers can be neglected, further coordination among the SSs is still needed such that theoretical guarantee is established for suppressing their possible interference to the other links.

\subsection{Tx-Beamforming with Complete Information about Interference Channels}
In this appendix we show that the global iterative power and joint transmit and receive beamformer converges when  each node has complete knowledge of the interfering channels towards other receivers. This provides a baseline for testing our algorithm and is brought for completeness. To that end, we introduce additional constraints on the ``interference leakage'' to the other links, i.e., ${R}_{n,i}\!=\!P_n\Vert\mathbf{u}^{\textrm{H}}_i\mathbf{H}_{n,i}\mathbf{w}_n\Vert^2$ (cf.~\cite{6189034}), as follows:
\begin{subequations}\label{eq_local_optimal_MSE_Tx_constrants}
\begin{align}
  \tag{\ref{eq_local_optimal_MSE_Tx_constrants}}
    \mathbf{w}^*_n &= \arg\min\limits_{\mathbf{w}_n} \; E\left\{\Vert\hat{s}_n-s_n\Vert^2\right\} \\
    \label{eq_local_optimal_MSE_Tx_constrants_a}
    & \textrm { s.t.}\; \;  \mathbf{w}^{\textrm{H}}_n\mathbf{w}_n \le P_n, \\
    \label{eq_local_optimal_MSE_Tx_constrants_b}
    &\; P_n\mathbf{u}^{\textrm{H}}_i\mathbf{H}_{n,i}\mathbf{w}_n\mathbf{w}^{\textrm{H}}_n\mathbf{H}^{\textrm{H}}_{n,i}\mathbf{u}_i\le {\tilde{R}_{n,i}}, \forall i\ne n,
\end{align}
\end{subequations}
where in (\ref{eq_local_optimal_MSE_Tx_constrants_b}) $\tilde{R}_{n,i}$ is the perceived interference from link $n$'s transmitter to link $i$'s receiver before the strategy update, such that the incurred interference to the other links is non-increasing when compared with the last decision round.
Writing the Lagrange dual problem of (\ref{eq_local_optimal_MSE_Tx_constrants}), the objective function can be expressed as
\begin{equation}
  \label{eq_modified_lagrange_objective}
  \begin{array}{ll}
      L(\mathbf{w}_n; \lambda_n, \kappa^n_{i\ne n})= \textrm{MSE}_n+ \lambda_n(\mathbf{w}^{\textrm{H}}_n\mathbf{w}_n-P_n) \\
      +\displaystyle\sum\limits_{i\ne n}\kappa^n_i\left(P_n\mathbf{u}^{\textrm{H}}_i\mathbf{H}_{n,i}\mathbf{w}_n\mathbf{w}^{\textrm{H}}_n\mathbf{H}^{\textrm{H}}_{n,i}\mathbf{u}_i - {\tilde{R}_{n,i}}\right),
  \end{array}
\end{equation}
where $\kappa^n_i$ is the Lagrange multiplier corresponding to (\ref{eq_local_optimal_MSE_Tx_constrants_b}) for the constraint on the interference to some other link $i$ ($i\in\mathcal{N}/\{n\}$).

Again, following the same technique of deriving (\ref{eq_local_optimum_TxBF_multiplier}), the receive filter based on (\ref{eq_mse_expand}) and (\ref{eq_modified_lagrange_objective}) is given by:
\begin{equation}
  \label{eq_local_optimum_TxBF_multiplier_modified}
    \mathbf{w}^*_n = \sqrt{P_n}\left(\sum\limits_{i=1}^N\kappa^n_i P_n \mathbf{H}^{\textrm{H}}_{n,i}\mathbf{u}_i\mathbf{u}^{\textrm{H}}_i\mathbf{H}_{n,i}
    + \lambda_n\mathbf{I}\right)^{-1}\mathbf{H}^{\textrm{H}}_{n,n}\mathbf{u}_n,
\end{equation}
where $\kappa^n_n=1$. The solution to (\ref{eq_local_optimal_MSE_Tx_constrants}) can be derived similarly to (\ref{eq_optimizing_TxBF}), i.e., with the Lagrange multipliers in (\ref{eq_local_optimum_TxBF_multiplier_modified}) computed using an off-the-shelf solver. By construction of the optimization problem in (\ref{eq_local_optimal_MSE_Tx_constrants}), we are able to show that the monotonicity of the power strategy sequence $\{P_i(t)\}_{t=0}^{\infty}$, $\forall i\in\mathcal{N}$ is guaranteed with Algorithm~\ref{alg_two_stage}. Obviously, the solution to (\ref{eq_local_optimal_MSE_Tx_constrants}) results in no SS increasing its interference to the other links by updating its Tx-beamformer. Then, with exactly the same technique of proving Theorem~\ref{thm_convergence_two_stage} based on the monotonically non-decreasing SINR (equivalently, non-increasing MMSE), we are able to ensure the convergence of Algorithm~\ref{alg_two_stage} using (\ref{eq_local_optimal_MSE_Tx_constrants}) for iteratively updating the Tx-beamformers:
\begin{Corollary}
  \label{Cor_2}
  Assume that the joint power allocation and Rx/Tx-beamforming problem as given in (\ref{eq_local_optimal_instance}) has a feasible solution. Then, with complete information about the perceived interference from every SS to each DS and the Rx-beamformer of each DS, Algorithm~\ref{alg_two_stage} with the iterative Tx-beamforming scheme based on (\ref{eq_local_optimal_MSE_Tx_constrants})  is guaranteed to converge.
\end{Corollary}

%\section*{Acknowledgment}
%The authors would like to thank the anonymous reviewers for their valuable comments and suggestions which improved the quality of the paper.
%\linespread{1.27}
\bibliographystyle{IEEEtran}
\bibliography{bibfile}

\end{document}